\documentclass[twocolumn]{aastex63}

\usepackage{amsmath}	
\usepackage{enumitem}
\usepackage[space]{grffile}  
\usepackage{bm} 

\def\Kepler{\textit{Kepler}} 
\def\TESS{\textit{TESS}} 

\defcitealias{2020AJ....160..276H}{H20}

\received{}
\revised{}
\accepted{}
\shorttitle{Impact of Planetary Companions on RV Follow-up of Transiting Planets}
\shortauthors{He, Ford, and Ragozzine}
\graphicspath{{./}{Figures/}}

\begin{document}

\title{Friends and Foes: Conditional Occurrence Rates of Exoplanet Companions and their Impact on Radial Velocity Follow-up Surveys}

\correspondingauthor{Matthias Yang He}
\email{myh7@psu.edu}

\author[0000-0002-5223-7945]{Matthias Y. He}
\affiliation{Department of Astronomy \& Astrophysics, 525 Davey Laboratory, The Pennsylvania State University, University Park, PA 16802, USA}
\affiliation{Center for Exoplanets \& Habitable Worlds, 525 Davey Laboratory, The Pennsylvania State University, University Park, PA 16802, USA}
\affiliation{Center for Astrostatistics, 525 Davey Laboratory, The Pennsylvania State University, University Park, PA 16802, USA}
\affiliation{Institute for Computational \& Data Sciences, 525 Davey Laboratory, The Pennsylvania State University, University Park, PA 16802, USA}

\author[0000-0001-6545-639X]{Eric B. Ford}
\affiliation{Department of Astronomy \& Astrophysics, 525 Davey Laboratory, The Pennsylvania State University, University Park, PA 16802, USA}
\affiliation{Center for Exoplanets \& Habitable Worlds, 525 Davey Laboratory, The Pennsylvania State University, University Park, PA 16802, USA}
\affiliation{Center for Astrostatistics, 525 Davey Laboratory, The Pennsylvania State University, University Park, PA 16802, USA}
\affiliation{Institute for Computational \& Data Sciences, 525 Davey Laboratory, The Pennsylvania State University, University Park, PA 16802, USA}
\affiliation{Institute for Advanced Study, 1 Einstein Drive, Princeton, NJ 08540, USA}
\affiliation{Center for Computational Astrophysics, Flatiron Institute, New York, NY 10010, USA}

\author[0000-0003-1080-9770]{Darin Ragozzine}
\affiliation{Department of Physics \& Astronomy, N283 ESC, Brigham Young University, Provo, UT 84602, USA}



\begin{abstract}

Population studies of \Kepler{}'s multi-planet systems have revealed a surprising degree of structure in their underlying architectures. Information from a detected transiting planet can be combined with a population model to make predictions about the presence and properties of additional planets in the system. Using a statistical model for the distribution of planetary systems \citep{2020AJ....160..276H}, we compute the \textit{conditional occurrence} of planets as a function of the period and radius of \Kepler{}--detectable planets.
About half ($0.52 \pm 0.03$) of the time, the detected planet is \textit{not} the planet with the largest semi-amplitude ($K$) in the system, so efforts to measure the mass of the transiting planet with radial velocity (RV) follow-up will have to contend with additional planetary signals in the data.
We simulate RV observations to show that assuming a single--planet model to measure the $K$ of the transiting planet often requires significantly more observations than in the ideal case with no additional planets, due to the systematic errors from unseen planet companions.
Our results show that planets around 10-day periods with $K$ close to the single--measurement RV precision ($\sigma_{1,\rm obs}$) typically require $\sim 100$ observations to measure their $K$ to within 20\% error.
For a next generation RV instrument achieving $\sigma_{1,\rm obs} = 10$ cm/s, about $\sim 200$ (600) observations are needed to measure the $K$ of a transiting Venus in a \Kepler{}--like system to better than 20\% (10\%) error, which is $\sim 2.3$ times as many as what would be necessary for a Venus without any planetary companions.

\end{abstract}

\keywords{Exoplanet systems (484); Exoplanet catalogs (488); Exoplanet detection methods (489); Exoplanets (498); Planet hosting stars (1242); Solar system planets (1260); Radial velocity (1332); Computational methods (1965)}


\section{Introduction} \label{sec:Intro}

Our current understanding of the inner regions of planetary systems around main sequence stars is largely shaped by detailed analyses of data obtained from the \Kepler{} primary mission \citep{2010Sci...327..977B, 2011ApJ...728..117B, 2011ApJ...736...19B, 2013ApJS..204...24B}. These analyses are enabled by advancements including, but not limited to, our models of the \Kepler{} detection efficiency \citep{2017ksci.rept...18C, 2019AJ....158..109H, 2020AJ....160..159C}, improved precision of the stellar properties \citep{2017AJ....154..108J, 2020AJ....159..280B}, and the development of novel statistical methods \citep{2018AJ....155..205H, 2019MNRAS.490.4575H}. In particular, the fully automated vetting of the \Kepler{} data allows for detailed forward modeling of planet occurrence rates as a function of orbital period and planet radius \citep{2019AJ....158..109H, 2020AJ....159..248K, 2021AJ....161...36B}. Even more powerful insights into the architectures of planetary systems can be gained from leveraging the information in \Kepler{}'s systems with multiple observed transiting planets \citep{2018AJ....156...24M, 2018ApJ...860..101Z, 2019MNRAS.490.4575H, 2020AJ....159..281G, 2020AJ....159..164Y}. Studies focusing on the properties of these multi-transiting systems have revealed a surprising degree of structure from system to system, including a tendency for planets with similar sizes and masses, preferential size orderings, clustered periods, uniform spacings, and an anti--correlation between mutual inclinations and planet multiplicity \citep{2013ApJ...763...41C, 2017ApJ...849L..33M, 2018AJ....155...48W, 2018ApJ...860..101Z, 2019MNRAS.490.4575H, 2020AJ....159..164Y, 2020AJ....159..281G, 2020AJ....160..276H}.
A common result is that the vast majority of planetary systems have multiple planets, often including planets that are not yet known.

\subsection{Finding and characterizing additional planets in known systems}

Just as detecting large populations of planets via systematic (e.g. transit) surveys have rapidly expanded our knowledge of the underlying distribution of planetary systems, this information can be used to make predictions about the presence of additional, non--transiting or otherwise yet--undetected planets in known systems. Previously, studies have typically relied on only using dynamical arguments to rule out regions of parameter space, estimating where additional planets \textit{could be} in a given system and still remain stable \citep{2009ApJ...699L..88R, 2015ApJ...807...44P, 2020AJ....159..188Z}.
Now, population models have become sophisticated enough that it is possible to predict where new planets are statistically \textit{likely to be} given the properties of the known planets and our knowledge of the system--level trends in their architectures. For example, \citet{2020AJ....160..107D} recently developed a framework for this purpose, using population models including one based on our clustered periods and sizes model \citep{2019MNRAS.490.4575H} and demonstrating the predictive utility applied to \TESS{} systems.

One important use of models that can infer the likely properties of as-yet unknown planets is for understanding what observational methods would be most beneficial for improved characterization. In particular, additional radial velocity (RV) measurements are especially synergistic with transiting detections for a number of reasons. First, the population models described above predict a significant fraction of yet--undetected planets in systems with a transiting planet.  Many of these planetary companions would be unseen in photometry due to the diminishing geometric transit probability with increasing orbital period, but would still be accessible with RV observations. 
Second, transit--like signals at the limits of the survey baseline (i.e. single transit events) are particularly tenuous yet valuable, as they are potentially caused by long--period planets -- their planetary nature can be confirmed by adequate RV follow-up \citep{2016AJ....152..206F, 2019AJ....157..248H, 2019AJ....157..218K}. This is especially common for the data gathered during the \TESS{} primary mission \citep{2015JATIS...1a4003R}, due to the relatively short baselines of each sector ($\sim 28$d) and uneven coverage for targets in different regions of the sky. Finally, sufficient RV measurements of a planetary system -- if adequately modeled -- can yield the mass of a given planet, a fundamental property which is typically unknown for most transiting planet candidates.
Indeed, a primary science goal of \TESS{} is to determine the masses of at least 50 planets smaller than $4 R_\oplus$ via RV follow-up of \TESS{} planet candidates. Planet masses from Transit Timing Variations (TTVs) are not expected to contribute significantly \citep{2019AJ....158..146H}.

\subsection{Planning Doppler exoplanet surveys}

The availability of ground-based telescopes with high--precision RV capabilities is a limiting factor in our ability to measure the masses of a large number of transiting planets with adequate precision. To date, over 2200 planet candidates have been discovered with \TESS{} \citep{2021ApJS..254...39G}, but only $\sim 120$ have been confirmed (and a subset of these with masses determined). Nearby stars with transiting planets detected from other surveys (e.g., \Kepler{} and K2) are also desirable targets for RV follow-up \citep{2019ApJS..244...11K, 2020SPIE11447E..42G}. Given so many promising targets (with even more on the way as \TESS{} data continues to be analyzed and obtained through its extended missions) and such precious telescope time, careful prioritization of the targets is crucial for maximizing the rate and quality of scientific discovery \citep{2019ESS.....430906Q}. Ideally, the ability to estimate the minimum number of RV observations required to achieve a certain goal (e.g., to measure the mass of a given planet to within a desired precision) would allow for the most efficient allocation of telescope time. These calculations may also be important for reducing (or at least, quantifying) the reporting bias that results from only publishing the easiest-to-measure and most well-constrained planet masses, which currently influences our derived planet mass--radius (M-R) relationships \citep{2018RNAAS...2...28M, 2018AJ....156...82C, 2020arXiv201111560T, 2021MNRAS.503.5504C}.

In the recent decade, there are significant efforts on multiple fronts to improve the precision of our RV measurements to new thresholds \citep{2016PASP..128f6001F}. From advancements in instrumental design and calibration, to correcting for telluric contamination and barycentric motion (i.e., the influence of Earth's atmosphere and velocity, respectively), astronomers have been able to achieve RV single--measurement precisions on the order of 1 m/s at a variety of observatories. The current and next generation spectrographs are promising sub--m/s RV precisions; the NASA commissioned NEID spectrograph \citep{2016SPIE.9908E..7HS} on the WIYN Telescope, the EXtreme PREcision Spectrograph (EXPRES; \citealt{2016SPIE.9908E..6TJ}) on the Lowell Discovery Telescope, and the Echelle SPectrograph for Rocky Exoplanets and Stable Spectroscopic Observations (ESPRESSO; \citealt{2021A&A...645A..96P}) on the Very Large Telescope are currently being tested and pushing 0.3 m/s or better precision. In addition to these technologies, there is active research in modeling stellar noise and applying novel statistical methods to extract the smallest signals from the data \citep{2017ApJ...846...59D, 2017arXiv171101318J, 2018A&A...620A..47D, 2021MNRAS.505.1699C, 2020A&A...633A..76C, 2020arXiv201100003D, 2020ApJ...905..155G, 2020MNRAS.492.3960R, 2020MNRAS.491.4131Z}.

Model comparison to quantify the evidence for multiple planets in a given set of RV observations is logically straightforward, but in practice is computationally difficult \citep{2007ASPC..371..189F, 2020AJ....159...73N}.
In cases where a complementary detection technique provides independent evidence for planets (e.g., transits, direct imaging, etc.), the problem shifts from model comparison to uncertainty quantification which is much easier computationally.
For example, in RV follow-up of planet candidates identified by transit surveys, the orbital period, phase and inclination are typically constrained more precisely by transit data than they would be after hundreds of RV observations.  Therefore, researchers often use the parameter values inferred from transit observations effectively as priors for the analysis of RV observations.
If one assumes a circular orbit (e.g., for short-period planets subject to tidal damping), then the shape of the RV perturbations is effectively known, leaving only the RV semi-amplitude $K$ (directly proportional to the planet mass) to be measured.  
For planets that may have a significant orbital eccentricity, there are two other parameters (eccentricity, $e$, and argument of pericenter, $\omega$) that affect the RV signature. 
Still, knowledge of the orbital period and phase dramatically simplifies the analysis, since the likelihood is often highly multimodal in period and phase, but is typically smooth and unimodal in $K$, $e \sin \omega$, and $e \cos \omega$.\footnote{There are often multiple modes for eccentricities pathologically close to unity, but these are rare and often can be excluded on other physical grounds, such as orbital stability or planet passing inside the host star's Roche limit at pericenter.}

One aspect of RV analysis that has not yet been studied in detail is how additional planets in a given system may affect the interpretation of the RV time series.
Ragozzine et al., submitted, note apparent differences in the architectures of planetary systems discovered by RVs and those seen in the \Kepler{} population.  They test whether this is due to the well-known challenges of clearly identifying and distinguishing multiple low-amplitude similar-period RV signals in these systems. They took known multi-planet systems from \Kepler{}, simulated RV observations under a variety of circumstances, and used a simple algorithm to determine which planets would have been discovered. They confirmed that simple signal-to-noise (SNR) metrics based on Gaussian statistics are not adequate for determining detectability since planets with highly significant signals were often missed, especially if they were close in period to other planets.

In addition to the challenges of identifying unknown planets, the presence of additional, unmodeled planets adds a source of systematic error that may affect the accuracy in the transiting planet's inferred mass (e.g., \citealt{2017A&A...602A..88A, 2017AJ....154..122C, 2017A&A...608A..35C, 2018A&A...618A.142B}).
If one ignores the possibility of additional undetected planets (i.e. model misspecification), then the formal measurement precision may be overly optimistic compared to the true measurement accuracy.
\citet{2018AJ....156...82C} modeled the observational effort required to characterize the planet yield predicted for the \TESS{} primary mission from \citet{2015ApJ...809...77S}, including potential unseen planets as a quadrature term in the effective RV uncertainty.
Identifying and characterizing exoplanets with RVs is aided significantly when one (or more) planets are known to transit. As RV follow-up of transiting planets is a common observing mode, we attempt to quantify the challenges resulting from unknown additional planets in these systems in this paper. We use the full power of our model for the distribution of inner planetary systems from \citet{2020AJ....160..276H} (hereafter \citetalias{2020AJ....160..276H}) which has been carefully tuned to infer the underlying distribution of \Kepler{} exoplanetary systems by matching multiple architectures--related metrics.

\subsection{Overview}

We organize this manuscript as follows. In \S\ref{sec:Methods}, we describe our methodology. We present the main results of our analyses in \S\ref{sec:Results}, beginning with calculations of \textit{conditional} planet occurrence rates and other statistics (\S\ref{results:PR_rates}--\ref{results:PR_Kmax}) before showing the distribution of RV observations required to measure the $K$'s of the transiting planets (\S\ref{results:RV_obs}--\ref{results:RV_precision}). In \S\ref{sec:Discussion}, we discuss implications for measuring the $K$'s of Venus--like planets (\S\ref{discussion:Venus}), the properties of unseen planets that have the largest impact on RV observations (\S\ref{discussion:Which_planets_affect_most}), as well as the limitations of this work including potential avenues for future improvements (\S\ref{discussion:Improvements}). Finally, we conclude and summarize our main takeaways in \S\ref{sec:Conclusions}, including implications for \TESS{} follow-up.

\section{Methods} \label{sec:Methods}

\subsection{General approach}

In \citetalias{2020AJ....160..276H} we presented an architectural model for the distribution of planetary systems by combining a parametric model (developed in \citealt{2019MNRAS.490.4575H, 2021AJ....161...16H}) and the principle of angular momentum deficit (AMD)-based stability \citep{2017A&A...605A..72L, 2017A&A...607A..35P}.
The AMD of a planetary system is the difference in its total angular momentum compared to a planetary system with the same set of masses and semi-major axes but coplanar and circular orbits. Thus, AMD measures how much ``extra" angular momentum there is in the eccentricities and inclinations of the planetary orbits, and is a nearly-conserved quantity even as this extra angular momentum is transferred between the planets due to typical gravitational interactions. The AMD stability criterion requires that the total AMD is insufficient to cause overlapping orbits or instabilities arising from mean-motion resonance overlap regardless of how it is partitioned amongst the individual planets.

This model, referred to as the ``maximum AMD model" in \citetalias{2020AJ....160..276H}, is the most recent in the suite of Planetary System Simulation models known as ``\textit{SysSim}''; we will simply refer to it as the ``\citetalias{2020AJ....160..276H} model" throughout this paper. Here, we use this model to compute the \textit{conditional} occurrence and properties of planets in systems with transiting planets. We address the following main questions:

\begin{itemize}
 \item Given a detected transiting planet with a certain period and radius, what is the distribution of other planets in such systems?
 \item How often does a transiting planet also have the largest RV semi-amplitude ($K$) if it is in a multi-planet system? What is the distribution of RV semi-amplitudes for planets in these systems?
 \item How many RV observations are required to measure the mass of a transiting planet with a certain precision, given the presence of additional planets in these systems?
\end{itemize}

In this section, we describe our procedure for analyzing these questions. We begin by briefly summarizing the \citetalias{2020AJ....160..276H} model and why it can be leveraged for this kind of analysis in \S\ref{methods:Models}. The methods are then described in two main parts: (\S\ref{methods:Conditioning}) how the simulated catalogs from the \citetalias{2020AJ....160..276H} model can be conditioned on to generate a large number of planetary systems with a given type of planet, and (\S\ref{methods:RV_obs}) how we simulate RV follow-up observations of such systems to quantify the impact of planetary companions on our ability to measure the signal of the conditioned planet. Likewise, the results in \S\ref{sec:Results} will be presented in a similar manner, with \S\ref{results:PR_rates}-\ref{results:PR_Kmax} relying on the methods in \S\ref{methods:Conditioning} only, and \S\ref{results:RV_obs}-\ref{results:RV_precision} involving the procedures in \S\ref{methods:RV_obs}.

\subsection{Models} \label{methods:Models}

First, we briefly summarize the relevant features of the \citetalias{2020AJ....160..276H} model to motivate why it is particularly powerful for probing the conditional statistics of planetary companions to transiting planets:

\begin{itemize}[leftmargin=*, label={}]
 \item \textbf{Planet clusters:} the properties of planets in multi-planet systems are correlated, with each system being composed of ``clusters" of planets with similar sizes and localized orbital periods (adequately spaced for stablility). In addition to the clustered periods and planet radii within each system, the distribution of periods and radii approximately follow power-laws for the population as a whole.
 \item \textbf{Mass-radius relation:} a probabilistic M-R relation is used to assign planet masses conditioned on their radii, with the non-parametric \citet{2018ApJ...869....5N} model used for radii above $1.47 R_\oplus$ and the ``Earth--like rocky" model from \citet{2019PNAS..116.9723Z} with a scatter that scales with radius for smaller sizes.
 \item \textbf{AMD stability and orbital architectures:} the critical AMD of each system is distributed amongst its planets to create dynamically packed yet AMD-stable orbital configurations. The eccentricities and mutual inclinations of the planets are automatically drawn this way, with higher multiplicity systems having lower average eccentricities and inclinations than systems with fewer planets. A minimum separation in mutual Hill radii between adjacent planets is also enforced.
\end{itemize}

Our model has been shown to reproduce several of the observed intra--system correlations described as the ``peas-in-a-pod'' features (e.g., \citealt{2017ApJ...849L..33M, 2018AJ....155...48W, 2020AJ....159..281G}), but does not capture more complicated trends. Namely, it provides an excellent description of the period ratio and transit depth ratio distributions for the \Kepler{} planet candidates, but underestimates the high degree of uniform spacings and preferential size orderings (the finding that larger planets tend to be exterior to smaller planets, to an extent greater than would be expected purely due to observational biases; see \citetalias{2020AJ....160..276H} and \citealt{2020AJ....159..281G} for more details). The model also does not account for the observed radius valley (e.g., \citealt{2017AJ....154..109F, 2018MNRAS.479.4786V}), a caveat which will be further discussed in \S\ref{discussion:Improvements}. Yet, the model provides an excellent fit to numerous additional constraints from \Kepler{} including but not limited to the overall rate of planets per star, variations in the frequency of planetary systems as a function of stellar color, the number of observed single and multi-planet systems at each multiplicity order, and the circular-normalized transit duration and transit duration ratio distributions (see \S2.4 of \citetalias{2020AJ....160..276H} for a full list of the observational constraints).

\subsection{Conditioning on transiting planets} \label{methods:Conditioning}

\begin{figure*}
\centering
\includegraphics[scale=0.85,trim={0.8cm 0.5cm 1cm 0.5cm},clip]{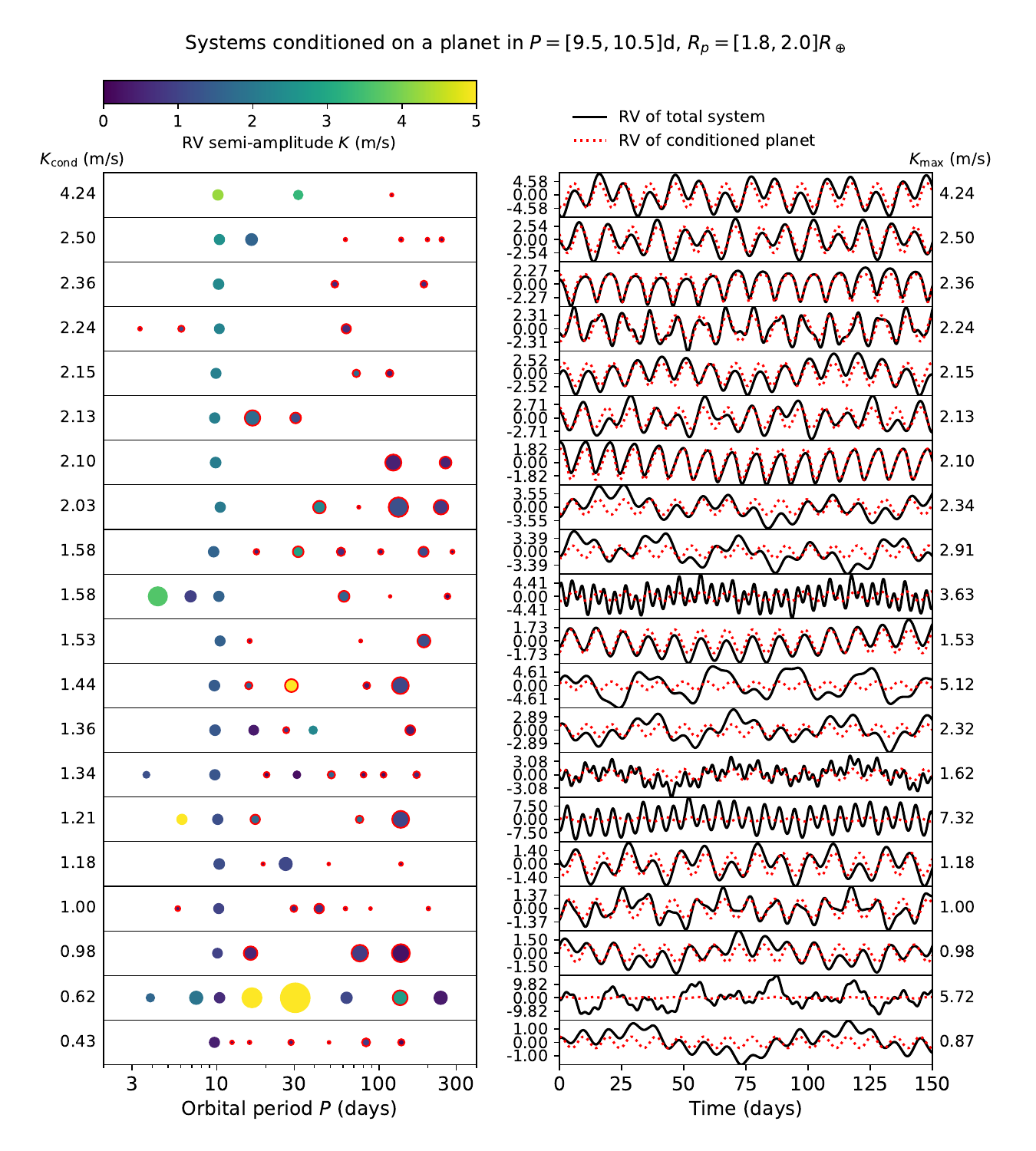}
\caption{Sample of 20 simulated systems from the \citetalias{2020AJ....160..276H} model, conditioned on an observed planet with $P_{\rm cond} = [9.5,10.5]$d and $R_{p,\rm cond} = [1.8,2.0]R_\oplus$. \textbf{Left--hand panel:} gallery of systems plotted along the orbital period axis. Circles with red outlines denote missed planets (non-transiting or otherwise undetected) while circles without outlines denote planets that would be observed by \Kepler{}; the size of each circle is proportional to the planet radius. Colors denote the RV semi-amplitude $K$ of each planet, as labelled by the color bar. \textbf{Right--hand panel:} RV time series of the systems on the left. Solid black curves denote the true RV signal of the full system, while dotted red curves denote the true RV signal of the conditioned planet only. The $K$'s of the conditioned planets ($K_{\rm cond}$) are listed on the left while the maximum $K$'s are listed on the right.}
\label{fig:systems_cond_1}
\end{figure*}

The ``\textit{SysSim}" codebase (\url{https://github.com/ExoJulia/SysSimExClusters}) allows for the generation of \textit{physical} and \textit{observed} catalog pairs. The former contains all of the planetary systems (including stellar properties) drawn directly from the model (as outlined in \S2.3.4 of \citetalias{2020AJ....160..276H}), while the latter contains only the observable properties of the transiting-and-\Kepler{}-detected planets simulated under a \Kepler{}-like primary mission. These observable properties include the measured period, transit depth, and transit duration of each detected planet, in addition to which system they belong to and the properties of the host star.

In this paper, we are interested in how knowledge of a (transiting) planet can be used to infer the distribution of additional planet companions conditioned on that planet, and how those additional planets affect our ability to measure the RV signal of that planet. Throughout this paper, we will also refer to the known planet (i.e., which is transiting and \Kepler{}--detectable, except for a subset of simulations in \S\ref{discussion:Venus}) as the ``conditioned planet" and likewise denote its parameters using subscripts ``cond"; for example, $P_{\rm cond}$, $R_{p,\rm cond}$, and $K_{\rm cond}$ refer to the period, radius, and RV semi-amplitude of the known planet. In order to condition on a given type of transiting planet, one can trivially select only the systems containing such a planet in a simulated observed catalog. The matching system in the physical catalog is then found, providing all of the planets (transiting or not) and their true properties. For example, one may wish to select all systems with a transiting planet in a given period and radius range. Because the physical catalog represents the true underlying distribution of planetary systems, the distribution of other planets in these systems can then be directly analyzed; in particular, we can also count how many planets would be missed by the (\Kepler{}) transit survey, compute the RV semi-amplitudes ($K$) of each planet, and evaluate how often the conditioned planet has the largest $K$ in the system.

To illustrate, in Figure \ref{fig:systems_cond_1} we show a sample of systems conditioned on a \Kepler{}-observed planet between $P_{\rm cond} = 9.5-10.5$ d and $R_{p,\rm cond} = 1.8-2 R_\oplus$. A gallery of the systems are plotted along orbital period in the left--hand panel; each circle represents one planet and is colored by the RV semi-amplitude $K$, with red outlines denoting planets that would be missed by that particular realization of \Kepler{}. The systems are sorted by $K_{\rm cond}$ (the $K$ of the conditioned planet), with larger amplitudes toward the top. Similar figures can be made using our code for any given range in $P_{\rm cond}$ and $R_{p,\rm cond}$, within the bounds of where our model is defined ($3-300$ d and $0.5-10 R_\oplus$).
These figures also serve as a check that our code for conditioning (and computing various conditional statistics) are working properly, and in particular that all conditioned systems actually contain at least one conditioned planet.

Depending on how many conditionals are chosen (i.e. how many criteria the planets must satisfy) and the range of desired values (e.g., how narrow the period and radius ranges are), one may require more than a single simulated (\Kepler{}-sized) catalog in order to acquire enough systems for a robust statistical analysis. Thus, for more involved calculations, we generate additional systems from our model until a desired number of transiting planets passing the conditionals is collected. We provide this additional functionality in the ``He\_Ford\_Ragozzine\_2021b'' branch of \textit{SysSim}, allowing users to simulate catalogs containing only systems with planets in a chosen period and radius range (and optionally, a planet mass range, and whether or not the given planet must also transit), instead of a statistical sample meant to match the \Kepler{} planet catalog.

\subsection{Simulating and fitting RV observations} \label{methods:RV_obs}

\subsubsection{The RV signal of a planetary system}

For each system, we first compute the theoretical radial velocity signal as a function of time assuming Keplerian motions of all the planets, following standard techniques. Ignoring interactions between the planets, the true RV signal $v_r(t)$ of the system at time $t$ is a sum of the RV signals of the individual planets, given by:
\begin{equation}
 v_r(t) = \sum_{i=1}^{n} K_i \Big[\cos{(\nu_i(t) + \omega_i)} + e_i \cos{(\omega_i)}\Big], \label{eq_rv}
\end{equation}
where $K_i$ is the semi-amplitude, $\nu_i$ is the true anomaly, $\omega_i$ is the argument of periapsis, and $e_i$ is the eccentricity, of the $i^{th}$ planet, and $n$ is the number of planets in the system. Here, only the true anomaly $\nu$ is a function of time, which for a given planet can be solved numerically in relation to the eccentric ($E$) and mean ($M$) anomaly:
\begin{eqnarray}
 \nu(t) = 2 \arctan{\bigg(\sqrt{\frac{1+e}{1-e}} \tan{(E(t)/2)}\bigg)}, \label{eq_true_anom} \\
 E(t) - e\sin{(E(t))} = M(t) \equiv \frac{2\pi(t - t_0)}{P}, \label{eq_ecc_mean_anom}
\end{eqnarray}
where $P$ is the orbital period and $t_0$ is a reference epoch which we draw uniformly between 0 and $P$.

The semi-amplitude $K$ in equation \ref{eq_rv} is commonly expressed in meters per second as \citep{2010exop.book...27L}:
\begin{equation}
 K = \frac{28.43 {\rm m/s}}{\sqrt{1-e^2}} \bigg(\frac{m_p\sin{i}}{M_{\rm Jup}}\bigg) \bigg(\frac{m_p + m_\star}{M_\odot}\bigg)^{-2/3} \bigg(\frac{P}{\rm yr}\bigg)^{-1/3}, \label{eq_K}
\end{equation}
where $m_p$ is the planet mass, $i$ is the orbital inclination relative to the sky plane, and $m_\star$ is the stellar mass. This is the key observable that can be inferred from RV measurements as it is directly related to the planet mass, which we wish to constrain. While RV observations alone only constrain the ``minimum" planet mass $m_p\sin{i}$ (the sky inclination is generally unknown from such observations alone, while the period and eccentricity can be fit from the RV time series), RV follow-up of transiting planets can yield the true planet mass since the inclination is constrained to a transiting configuration (i.e. $i \sim 90^\circ$).

\subsubsection{Measuring $K$ of the transiting planet}

We perform a suite of simulated RV observations and fitting in order to assess how many observations ($N_{\rm obs}$) are needed to accurately measure the semi-amplitude $K$ of the transiting planet (and thus its mass). We construct a series of RV measurements by adding daily observations with a small variation $\delta{t} \sim \mathcal{N}(0, \sigma_t = 0.2 {\rm days})$ in the exact time to avoid strong aliasing.\footnote{We also simulate a set of RV observations with an equivalent $N_{\rm obs}$, but randomly sampled in time over the same total duration, and verify that the results are similar. We default to the daily-added observations, given the nightly cadence of ground-based RV observations.} For each observation, we add an overall error term $\delta{v_r} \sim \mathcal{N}(0, \sigma_{1,\rm obs})$ to the true RV signal given by Equation \ref{eq_rv}, where $\sigma_{1,\rm obs}$ is the single--measurement precision. We adopt a nominal value of $\sigma_{1,\rm obs} = 0.3$ m/s but also test prospects of greater precisions as enabled by next generation extreme--precision RV instruments ($\sigma_{1,\rm obs} = 0.1$ m/s), and more common capabilities ($\sigma_{1,\rm obs} = 1$ m/s). We note that we do not explicitly include other known and potentially serious sources of systematic uncertainties such as those arising from stellar activity, and thus our recommendations should be viewed as optimistic.

For each set of RV observations, we fit the data points using a single--planet model (i.e., for the transiting planet) using generalized least squares (GLS) in order to measure $K_{\rm cond}$ (i.e. $K$ of the transiting planet).
The single--planet model assumes that the measured RVs can be modeled by a single term in the summation of Equation \ref{eq_rv} corresponding to the conditioned planet, which is a linear function of the vector $\bm{X} \equiv (X_1, \dots, X_{N_{\rm obs}})$:
\begin{equation}
 \bm{v_{r,\rm obs}} = K_{\rm cond}\bm{X} + \bm{\delta{v_r}}. \label{eq_rv_model}
\end{equation}
where $X_i = \cos(\nu_{\rm cond}(t_i) + \omega_{\rm cond}) + e\cos(\omega_{\rm cond})$. Simply put, $\bm{v_{r,\rm obs}}$ is the vector of RV observations at a vector of observation times $\bm{t}$. It can be shown that the GLS method estimates the best fitting $K_{\rm cond}$ by minimizing the length of the residual vector, giving:
\begin{equation}
 \hat{K}_{\rm cond} = (\bm{X}^T \bm{\Omega}^{-1} \bm{X})^{-1} \bm{X}^T \bm{\Omega}^{-1} \bm{v_{r,\rm obs}}, \label{eq_GLS_estimator}
\end{equation}
where $\bm{\Omega} = \bm{I} \sigma_{1,\rm obs}^2$ is the covariance matrix which we assume is diagonal.
In the calculation of $\bm{X}$, we also assume that the orbit of the transiting planet is known exactly. The orbital period and phase are typically measured with high precision by \Kepler{}, while the orbital eccentricity is typically not well constrained from transit data.  Nevertheless, we assume that its orbital eccentricity and pericenter direction are known (i.e., by adopting the value drawn in the physical catalog) for computational convenience. This can be viewed as slightly optimistic, since a real survey would need to account for covariance of $K$ with the eccentricity and pericenter direction.

The above procedure is repeated $n = 100$ times for each value of $N_{\rm obs}$. For most simulations, we vary the number of observations $N_{\rm obs}$ between 5 and 300 (with 20 uniformly--spaced points in log-space, rounded to the nearest integer); for more difficult to measure planets, we test up to $N_{\rm obs} = 10^3$. Finally, the minimum $N_{\rm obs}$ required to achieve a root-mean-square-deviation (RMSD) from the true $K_{\rm cond}$, of less than 20\%, i.e.:
\begin{equation}
 {\rm RMSD}(K_{\rm cond}) = \sqrt{\sum{(\hat{K}_{\rm cond} - K_{\rm cond})^2}/n} < 0.2 K_{\rm cond}, \label{eq_rmsd_Kcond}
\end{equation}
is determined for each system.

Our choice of computing the $N_{\rm obs}$ required for measuring $K_{\rm cond}$ to within 20\% of the true value is motivated by \citet{2019ApJ...885L..25B}, who recommend a $\pm 20\%$ mass precision for detailed atmospheric analyses of transiting planets based on simulations of JWST transmission spectra for most (i.e. terrestrial) planets. We also consider an accuracy of 10\% for some cases (\S\ref{discussion:Venus}). However, we emphasize that we choose to work with errors in measuring $K$, instead of directly in planet mass, throughout this paper. The uncertainty of inferring a planet's mass is also a function of the uncertainties in the other parameters in equation \ref{eq_K} (most notably orbital eccentricity and stellar mass). In particular, stellar mass uncertainties are often on the order of $\sim 7\%$, but can be much higher \citep{2020AJ....159..280B}.
Thus, an accuracy of 20\% in planet mass would likely require even more observations than what we compute in this study.

\subsubsection{Ideal (single--planet) case} \label{methods:Ideal_case}

To quantify how strongly the presence of other planets affects our ability to measure $K_{\rm cond}$, we also test how many RV observations are typically necessary under our assumptions if the transiting planet is indeed the only planet in the system. This is the ideal case in which the single--planet model used to fit the RV data is the correct model. We note that some simulated systems are true intrinsic singles. \textit{A priori}, one does not know how many additional planets (with significantly large $K$'s to affect the total RV signal) are in a given system; thus, a direct comparison between the minimum $N_{\rm obs}$ required in this ideal case (where all other planets are removed) versus the minimum $N_{\rm obs}$ required in the general case (where the conditioned systems include those with and without additional planets), to achieve the same precision in $K_{\rm cond}$, serves to determine the importance of accounting for multiplicity when planning observations or analyzing data from RV follow-up campaigns.

To check that our RV simulations and procedure outlined in this section are working as intended, we visually inspected a series of RV models with the fitted $K$'s for each $N_{\rm obs}$ plotted on top of the synthetic RV observations for a sample of conditioned systems. We verified that the accuracy in $K$ improves with $N_{\rm obs}$, and that the minimum $N_{\rm obs}$ decreases as $K_{\rm cond}/\sigma_{1,\rm obs}$ increases, as expected.

\section{Results} \label{sec:Results}

We organize our key results into two themes as follows. First, in \S\ref{results:PR_rates}-\ref{results:PR_Kmax} we condition on transiting planets to show how knowledge of a given planet shapes the distribution of other planets in the same system; the procedure of conditioning is described in \S\ref{methods:Conditioning}. In particular, we compute the planet occurrence rates as a function of period and radius conditioned on an example type of planet (\S\ref{results:PR_rates}), the conditional mean number of planets per system as a function of the conditioned planet's period and radius (\S\ref{results:PR_numpl}), and the fraction of times when there is another planet with a greater RV semi-amplitude than that of the conditioned planet (\S\ref{results:PR_Kmax}).

Then, in \S\ref{results:RV_obs}-\ref{results:RV_precision} we report our results for the number of RV observations ($N_{\rm obs}$) required to measure a given planet's semi-amplitude ($K_{\rm cond}$), following the procedure of simulating RV follow-up surveys as described in \S\ref{methods:RV_obs}. These results are explored in the following order: the distribution of $N_{\rm obs}$ needed for measuring the $K$'s of relatively short period planets (e.g. $8-12$ d, such as those discovered by \TESS{}) of various sizes (\S\ref{results:RV_obs}), how $N_{\rm obs}$ varies as a function of the conditioned planet's period and radius (\S\ref{results:RV_obs_PR}), and finally the effect of instrumental RV precision on $N_{\rm obs}$ (\S\ref{results:RV_precision}).

\subsection{Conditional occurrence rates of planetary companions} \label{results:PR_rates}

\begin{figure*}
\centering
\includegraphics[scale=0.45,trim={0.8cm 0 0 0.2cm},clip]{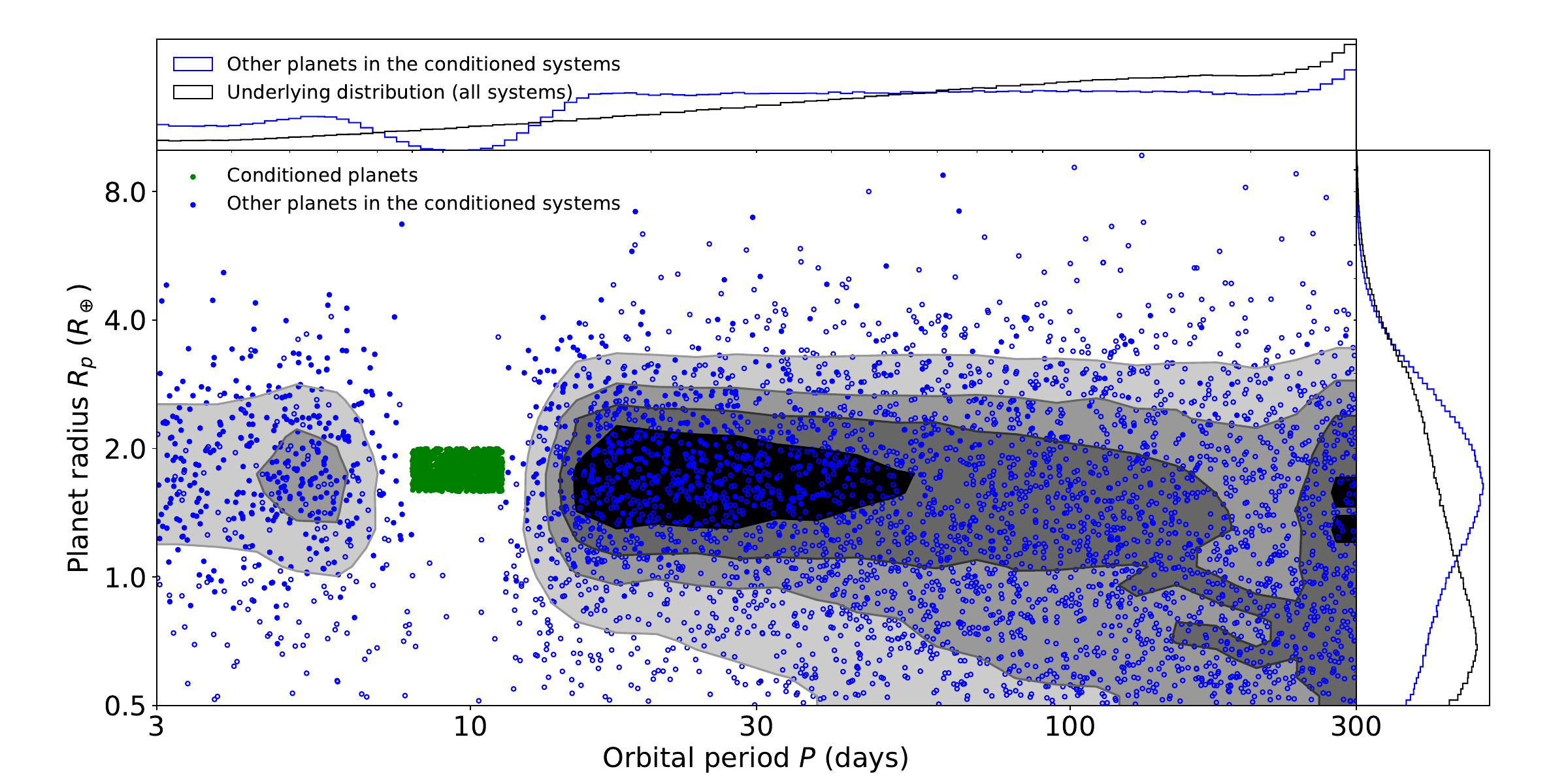}
\includegraphics[scale=0.45,trim={0.8cm 0 0 0.2cm},clip]{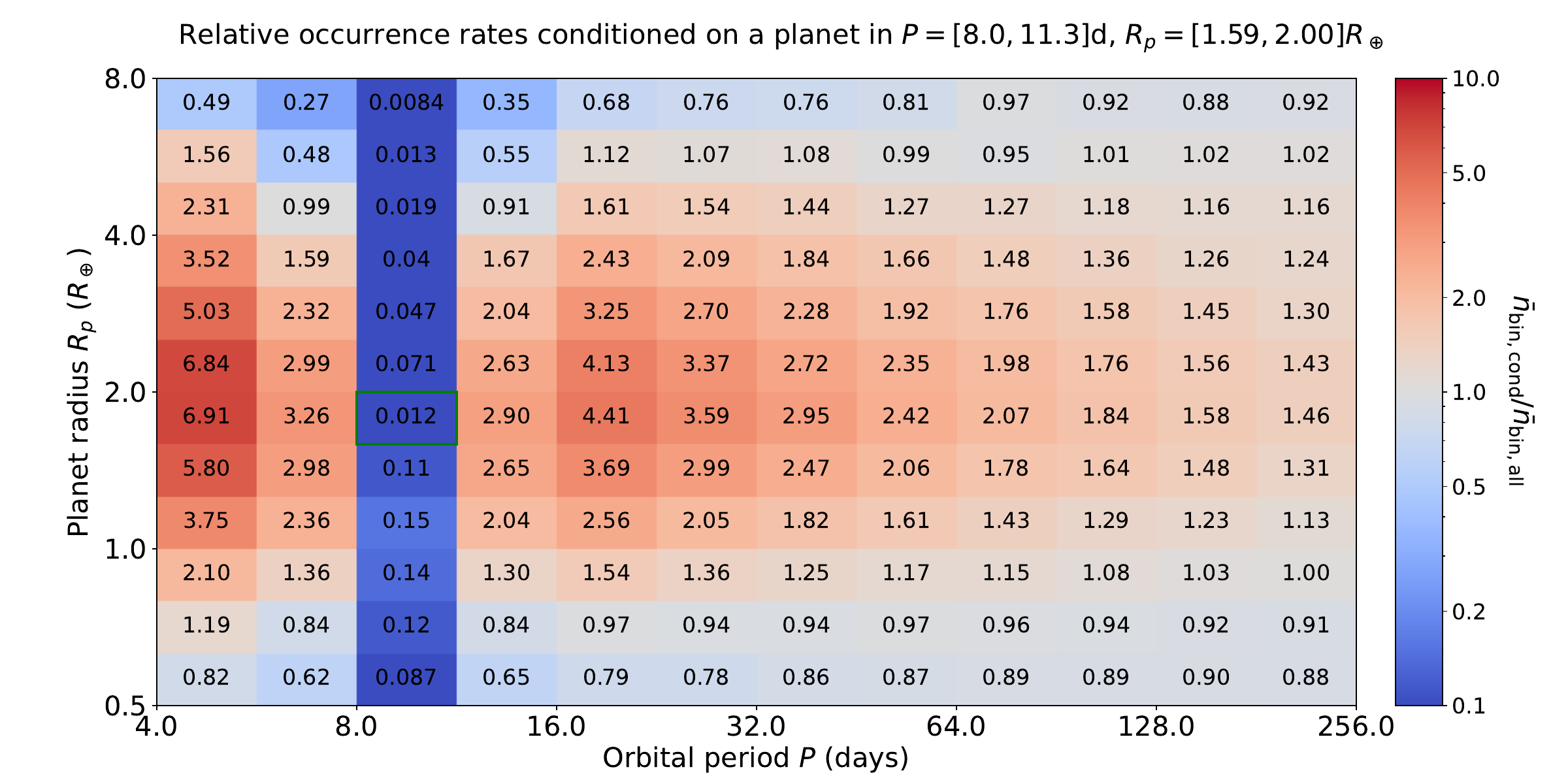}
\caption{\textbf{Top panel:} period--radius distribution of planets conditioned on a \Kepler{}--observed planet with $P_{\rm cond} \simeq [8,11.3]$ d and $R_{p,\rm cond} \simeq [1.59,2.0]R_\oplus$. Green and blue circles denote conditioned and other planets, respectively, in 1000 systems with a conditioned planet. Filled and hollow circles denote \Kepler{}--detected and undetected planets in the simulation, respectively. The shaded contours enclosing 11.8\%, 39.3\%, 67.5\%, and 86.5\% (innermost to outermost) of the points denote the distribution of planets for a much larger sample ($\sim 2.4\times10^5$) of conditioned systems. The marginal distributions of these systems are also shown as blue histograms on the top and side panels. For comparison, the black histograms show the marginal, underlying distributions for all systems (i.e. the full ensemble of planets in the \citetalias{2020AJ....160..276H} model).
\textbf{Bottom panel:} planet occurrence rates (intrinsic numbers of planets per star in each period--radius bin) of the conditioned systems relative to the occurrence rates of all systems. In other words, the number in each cell (also the color-scale) is the mean number of planets per star for the conditioned systems ($\bar{n}_{\rm bin,cond}$) divided by the mean number of planets per star over all systems ($\bar{n}_{\rm bin,all}$).
In bins overlapping with the conditioned period--radius range (marked by the green box), the conditioned planets themselves are excluded from the calculation (the value is nonzero because of the bin width in period, which occasionally allows two conditioned planets in the same system).}
\label{fig:PR_grid_cond_rates}
\end{figure*}

Analyses of planet occurrence rates from transit surveys commonly involve computing the \textit{mean number of planets per star} along a grid of orbital period $P$ versus planet radius $R_p$. The $P$-$R_p$ parameter space is chosen because these are closely related to the observables of a transit survey. Additionally, they provide an easily interpretable picture of how the frequency of planets varies as a function of their sizes and separations from their host stars. A natural extension of this type of calculation is possible using our models -- which characterize planetary systems, not just individual planets -- to explore how the occurrence of planets varies given knowledge of a transiting planet.

To illustrate the predictions of our model in such an example (where a transiting planet is known), we generate a large number of simulated planetary systems with a \Kepler{}--detected planet in $P_{\rm cond} \simeq 8 - 11.3$ d and $R_p \simeq 1.59 - 2 R_\oplus$. This period and radius range was chosen to resemble that of a ``typical'' super--Earth to sub--Neptune sized planet detected by \Kepler{} or \TESS{} (with the exact bounds selected to match a bin on a $12\times12$ grid of log--uniformly spaced $P$-$R_p$ bins as described below); the same procedure can be repeated over any $P_{\rm cond}$ and $R_{p,\rm cond}$ range covered by our model. In Figure \ref{fig:PR_grid_cond_rates} (top panel), we plot the distribution of planets in these systems. A sample of 1000 conditioned systems are shown as scatter points for individual planets, while the contours denote the 11.8\%, 39.3\%, 67.5\%, and 86.5\% regions of points over a larger set ($\sim 2.4\times10^5$) of conditioned systems. The histograms on the top and right sides show the marginal distributions of $P$ and $R_p$, respectively, for the other planets in the conditioned systems (blue) and for all planetary systems (black). The valley in the period distribution around $P_{\rm cond}$ is clearly sculpted by the mutual Hill stability criteria in our models, preventing other planets from being too close to the orbit of the conditioned planet (there is an exceedingly rare overlap between the blue and green points due to the width of the $P_{\rm cond}$ range). Yet, the occurrence of planets is enhanced at a close proximity both interior and exterior to the conditioned planet because of the clustering in orbital period for planets drawn in the same cluster. Likewise, the peak of the conditional planet radius distribution also shifts towards the size of the conditioned planet.

In the bottom panel of Figure \ref{fig:PR_grid_cond_rates}, we show the \textit{relative} occurrence rates on a $P$-$R_p$ grid, conditioned on the same set of systems. Here, we define the ``relative occurrence rate'' as the ratio of the intrinsic mean number of planets per star (in a given $P$-$R_p$ bin) in systems where \Kepler{} would detect a planet in the conditioned range, relative to the intrinsic mean number of planets per star (in the same $P$-$R_p$ bin) over all systems (i.e., without conditioning on the presence of a planet).
In each bin, we compute the conditional occurrence rate ($\bar{n}_{\rm bin,cond}$) and the overall occurrence rate ($\bar{n}_{\rm bin,all}$) of the model. The ratio of these two gives the relative occurrence rate for the conditioned systems, as shown by the numbers and the color-scale.
Consistent with the top panel, we find that the relative occurrence of planets is sharply diminished in bins sharing the same period range as $P_{\rm cond}$ due to orbital stability considerations, while it is most significantly boosted in bins with similar radii and periods roughly a bin-width away (i.e. a period ratio of $\sim \sqrt{2}$) both interior and exterior to the conditioned bin (marked by the green box).
We caution that while the \citetalias{2020AJ....160..276H} model was designed to capture correlations in multi-planet systems, it relies on power--law parametrizations of the overall period and radius distributions and is thus not as flexible as models from other occurrence rate studies (e.g. \citealt{2018AJ....155..205H, 2019AJ....158..109H}) in terms of the joint period--radius distribution (for reasons further described in \S\ref{discussion:Improvements}). However, computing conditional occurrence rates relative to the overall rates of the full model serves to quantify how conditioning on a known planet shifts the expected distribution of other planets in the same system.

\subsection{Conditional number of RV--detectable planets} \label{results:PR_numpl}

\begin{figure*}
\centering
\includegraphics[scale=0.45,trim={0.5cm 0 0.5cm 0.2cm},clip]{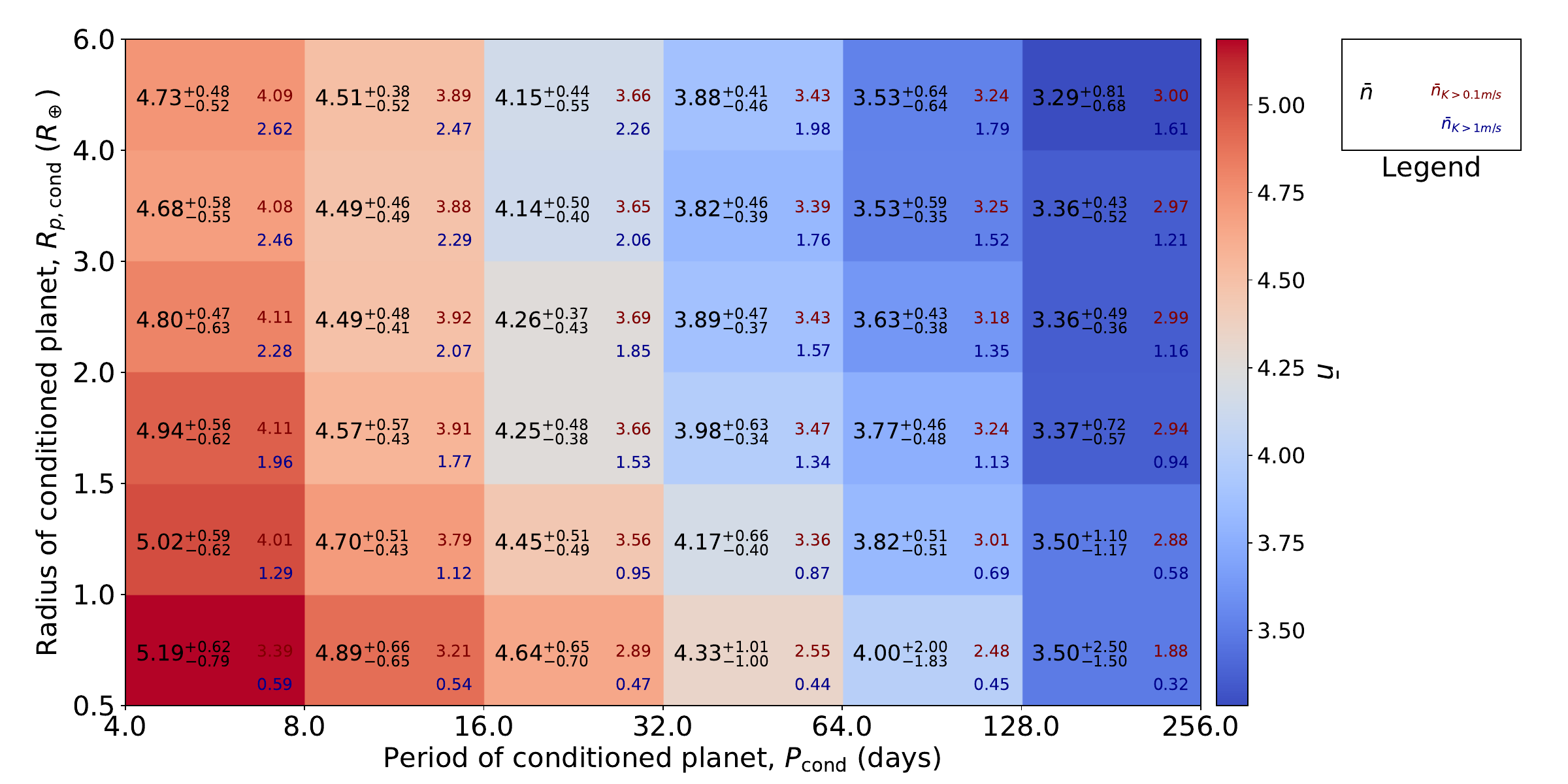}
\caption{Intrinsic numbers of planets in each system, conditioned on an observed planet for each period--radius bin. The numbers in each cell (as listed in the legend) are the mean intrinsic number of planets counting all planets ($\bar{n}$), planets with $K > 0.1$ m/s ($\bar{n}_{K > 0.1m/s}$), and planets with $K > 1$ m/s ($\bar{n}_{K > 1m/s}$). The centered values ($\bar{n}$) are also denoted by the color scale.}
\label{fig:PR_grid_mult}
\end{figure*}

In the previous section, we quantified the $P$-$R_p$ distribution of additional planets in systems conditioned on a given type of planet, and how that varies relative to all planets produced in our model in general. Here, we consider how the expected number of planets per system varies as a function of the period and radius of the conditioned planets (i.e. $P_{\rm cond}$ and $R_{p,\rm cond}$), and how many of those planets have significant $K$'s.

Using an ensemble of 100 \Kepler{}-sized simulated catalogs, we condition on systems with a \Kepler{}-detected planet in each $P_{\rm cond}$-$R_{p,\rm cond}$ bin, and count the intrinsic mean number of planets in each system. We note that unlike in Figure \ref{fig:PR_grid_cond_rates}, only the conditioned planet must be in the bin; the other planets in the system can have any size or period within the extent of our model.
In each bin of Figure \ref{fig:PR_grid_mult}, the black centered number (with uncertainties) denotes the mean number of planets including all planets in such systems ($\bar{n}$) -- this is also represented by the color scale. The uncertainties show the 68\% credible interval computed from 100 separate simulated catalogs. We also report two additional numbers, for the mean number of planets including just planets with $K > 0.1$ m/s (maroon) and $K > 1$ m/s (dark blue), including the conditioned planet.

Several conclusions can be drawn from this figure. First, $\bar{n}$ is significantly greater than one across all bins considered, indicating that the detection of a single transiting planet often implies the existence of multiple additional planets (within $0.5-10 R_\oplus$ and $3-300$ d) in the same system.
This is a reflection of the findings in previous studies (e.g. \citetalias{2020AJ....160..276H}) that most \Kepler{} planets are in multi-planet systems.
We note that we exclude bins for planets above $6 R_\oplus$ at all periods (e.g., hot Jupiters) because our model is not well--constrained for the population of larger planets and they are of relatively less interest for our purposes; to the latter point, hot Jupiters are typically single in inner planetary systems \citep{2012PNAS..109.7982S}.
While the mean number of planets per planetary system in this model is $3.12_{-0.28}^{+0.36}$ \citepalias{2020AJ....160..276H}, systems where transiting planets are detected tend to have higher than average multiplicities, since having more planets increases the probability of detecting at least one planet. Most of these planets also have RV semi-amplitudes greater than 0.1 m/s, although typically only one or two planets exhibit $K > 1$ m/s. The value of $\bar{n}$ decreases as the period of the conditional planet increases. There is also a slight gradient with increasing $\bar{n}$ towards smaller radii, likely due to the clustered nature of planet sizes; the detection of a small transiting planet suggests that there are additional similarly sized planets in the same system. The fact that $\bar{n}_{K > 0.1 {\rm m/s}}$ (and $\bar{n}_{K > 1 {\rm m/s}}$) show the opposite trend (i.e. decrease towards smaller radii) further supports this interpretation.

\begin{figure*}
\centering
\includegraphics[scale=0.45,trim={0.5cm 0 0.5cm 0.2cm},clip]{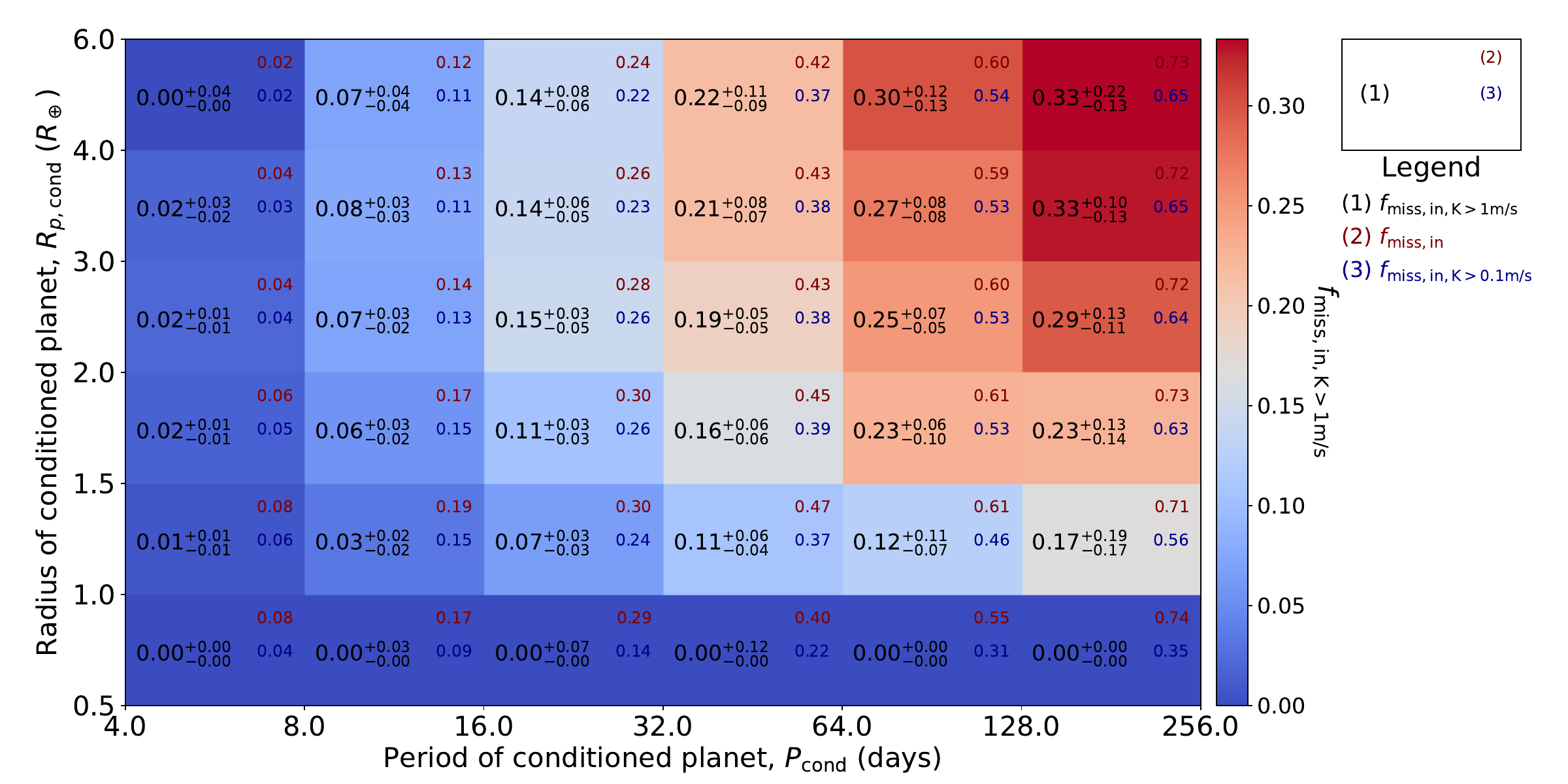}
\caption{Fractions of systems where \Kepler{} would have missed an interior planet, conditioned on an observed planet for each period--radius bin. The numbers in each cell (as listed in the legend) are the mean fractions of systems where: any interior planet is missed ($f_{\rm miss,in}$), an interior planet with $K > 0.1$ m/s is missed ($f_{\rm miss,in,K>0.1m/s}$), and an interior planet with $K > 1$ m/s is missed ($f_{\rm miss,in,K>1m/s}$). For example, $f_{\rm miss,in,K>1m/s} \equiv N_{\rm miss,in,K>1m/s}/N_{\rm tot}$ where $N_{\rm tot}$ is the number of systems with a conditioned planet in a given bin and $N_{\rm miss,in,K>1m/s}$ is the number of those systems where there is a missed planet with $K > 1$ m/s interior to the conditioned planet. The centered numbers with uncertainties show the 16\% and 84\% quantiles, which are also denoted by the color scale.}
\label{fig:PR_grid_fmissin}
\end{figure*}

In Figure \ref{fig:PR_grid_fmissin}, we show the fraction of observed systems where \Kepler{} would have missed an interior planet (transiting or not) conditioned on a detected planet in each $P$-$R_p$ bin. The black centered number (along with the color scale) and uncertainties denotes the fraction of the time where an interior planet with $K > 1$ m/s is missed; maroon and dark blue numbers include all missed--interior planets and just those with $K > 0.1$ m/s, respectively.\footnote{We note that our models only include planets between $3-300$ d, and thus ignore any planets less than 3 days (e.g. any ultra-short period planets). For this reason, the true fraction of missed interior planets is likely even higher than what we compute here.} The fraction is only a few percent for conditional planets in the smallest period bin ($4-8$ d) and increases toward longer periods, as expected since more planets can fit within wider orbits. Nevertheless, a significant fraction of these systems (as high as $30-40$\%) host interior planets with $K > 1$ m/s that would be undetected by a \Kepler{}-like transit survey! These results indicate the value of RV follow-up of transiting planets for finding additional short-period companions. Finally, the fraction of systems with missed--interior planets decreases as the size of the conditioned planet decreases. Due to the near coplanar nature of high multiplicity systems, most planets at shorter periods would likely be transiting and detectable if the planet hardest to detect (i.e. a small planet at a longer period) is already recoverable by transits. In other words, the discovery of a small planet often indicates a complete knowledge of the inner planetary system, while the discovery of a large planet (at longer periods) may simply mean a lack of sensitivity to smaller, interior planets.

\subsection{Frequency of planetary companions with more dominant RVs} \label{results:PR_Kmax}

\begin{figure*}
\centering
\includegraphics[scale=0.45,trim={0.5cm 0 0.5cm 0.2cm},clip]{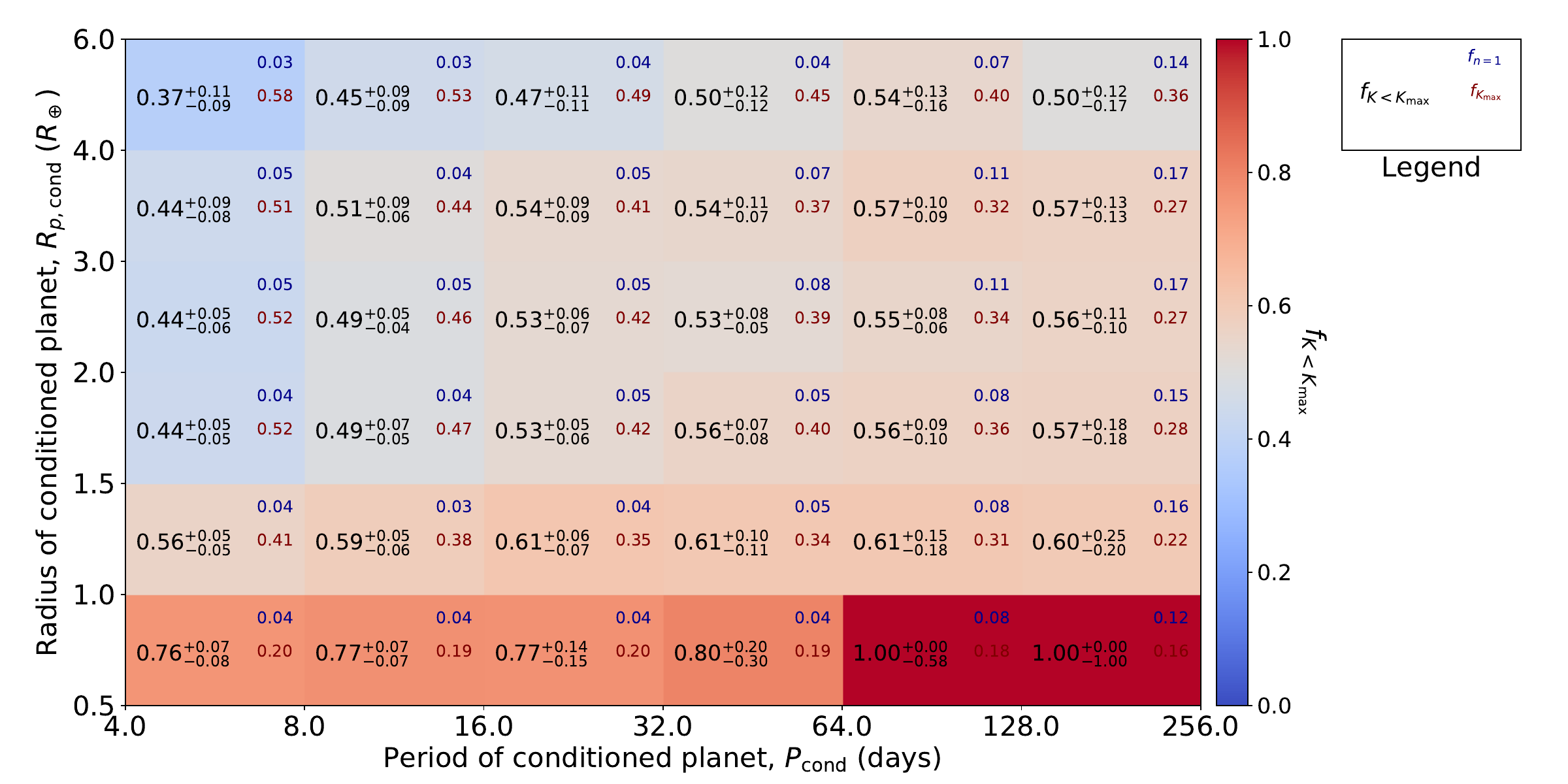}
\caption{Fractions of systems where another planet has a larger RV semi-amplitude $K$ compared to an observed planet for each period--radius bin. The numbers in each cell (as listed in the legend) are the (mean) fraction of systems where the observed planet: is the only planet in the system ($f_{n=1}$; i.e. intrinsic singles) and thus has the largest $K$ by default, \textbf{is in a multi-planet system and} has the largest $K$ ($f_{K_{\rm max}}$), or does not have the largest $K$ ($f_{K < K_{\rm max}}$; centered numbers with uncertainties denoting the 16\% and 84\% quantiles, which are also denoted by the color scale). In other words, $f_{K < K_{\rm max}} \equiv N_{K < K_{\rm max}}/N_{\rm tot}$ denotes the fraction of observed systems where another planet in the system dominates the RV signal. We note that the numbers in cells toward the bottom--right corner are dominated by small number statistics (due to the difficulty of transit detections for small planets at longer periods).}
\label{fig:PR_grid_fKmax}
\end{figure*}

Here, we consider how the RV semi-amplitude of the transiting planet compares to those of other planets in the same system. The fraction of observed systems where another planet in the system has a greater $K$ than the conditioned planet, $f_{K<K_{\rm max}} \equiv N_{K<K_{\rm max}}/N_{\rm tot}$, is shown in Figure \ref{fig:PR_grid_fKmax} (black centered numbers and color scale). We also compute the fraction of observed systems where the conditioned planet is the only planet in the system (blue numbers) and where the conditioned planet has the largest $K$ of all the planets in the system (red numbers). The sum of these three rates is 1 for each bin. Across most bins, the conditioned planet does not have the largest $K$ roughly half of the time. This fraction ($f_{K<K_{\rm max}}$) increases towards smaller planet radii as expected since a small planet tends to be less massive and thus its RV signal is likely dwarfed by that of another planet in the system. There is also a modest increase in $f_{K<K_{\rm max}}$ as the period of the conditioned planet increases, due to the weak dependence of $K \propto P^{-1/3}$. These results imply that efforts to follow-up planets of any size and period found by transit surveys (especially smaller planets at longer periods) with RV observations will often have to contend with additional and substantial Keplerian signals in the data that are not due to the transiting planet! In the next section, we show how these additional planets may affect the number of RV observations necessary to accurately measure $K_{\rm cond}$.

\subsection{Number of RV observations required for robust mass measurements of transiting planets} \label{results:RV_obs}

\begin{figure*}
\centering
\includegraphics[scale=0.75,trim={0.6cm 0 0.6cm 0.5cm},clip]{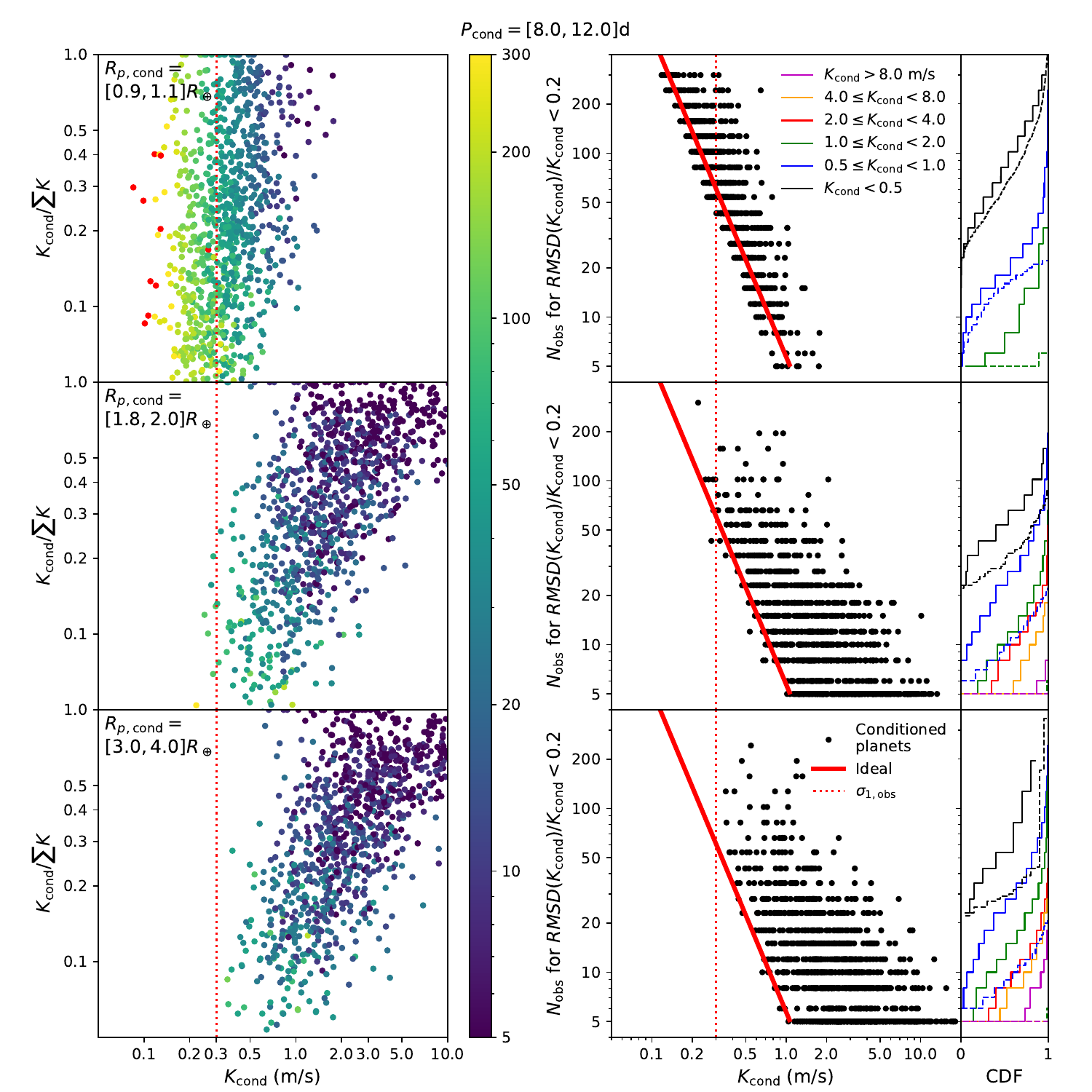}
\caption{Number of RV observations ($N_{\rm obs}$) required to measure $K_{\rm cond}$ to within 20\% accuracy, for three size regimes of the conditioned planet ($R_{p,\rm cond}$) as labeled: $[0.9,1.1] R_\oplus$ (top panels), $[1.8,2.0] R_\oplus$ (middle panels), and $[3.0,4.0] R_\oplus$ (bottom panels). All panels consider the same period range for the conditioned planet: $P_{\rm cond} = [8,12]$ d.
\textbf{Left--hand panels:} scatter plots of $K_{\rm cond}/\sum{K}$ vs. $K_{\rm cond}$ for the conditioned planet, colored by $N_{\rm obs}$. Points towards the top of the plot indicate conditioned planets that dominate the RV signal of its system (points with exactly $K_{\rm cond}/\sum{K} = 1$ are intrinsic singles). Points off the colorbar (red) indicate planets where more than $N_{\rm obs} = 300$ observations are needed to measure $K_{\rm cond}$. In all panels, the vertical dashed line denotes the single--measurement precision, $\sigma_{1,\rm obs} = 0.3$ m/s.
\textbf{Right--hand panels:} scatter plots of $N_{\rm obs}$ vs. $K_{\rm cond}$ for the same simulations in the left--hand panels. The solid red line shows a fit to the ideal single--planet case. The attached panels to the right show the CDFs of $N_{\rm obs}$ for points in each $K_{\rm cond}$ bin as labeled by the legend; solid lines represent the CDFs of the black points while dashed lines represent the analogous CDFs for the ideal case (i.e. $N_{\rm obs}$ predicted from the red line given the same distribution of $K_{\rm cond}$ as the black points). Note that some dashed lines may overlap (e.g. those rising to unity at $N_{\rm obs} = 5$).}
\label{fig:Kcond_Nobs_Rp}
\end{figure*}

A primary purpose of RV follow-up for transiting planets is to measure their masses. These measurements are particularly valuable because they allow us to constrain the bulk compositions of the planets in combination with their radii already inferred from the transits alone. However, robust mass measurements require adequately fitting the semi-amplitude $K$ of the planet from a number of RV observations limited by telescope time. We outlined our procedure for simulating a series of RV observations in \S\ref{methods:RV_obs} in order to determine the minimum number of data points necessary to constrain $K_{\rm cond}$. Here, we show the results of such simulations as applied to several cases of conditioned planets.

In Figure \ref{fig:Kcond_Nobs_Rp}, we consider three planet size regimes of interest: (1) Earth--sized planets ($R_{p,\rm cond} = [0.9, 1.1] R_\oplus$), (2) ``radius valley'' super--Earths ($R_{p,\rm cond} = [1.8, 2.0] R_\oplus$), and (3) Neptune--sized planets ($R_{p,\rm cond} = [3.0, 4.0] R_\oplus$). To keep other variables constant, we use the same period range, $P_{\rm cond} = [8, 12]$ d, for all three cases. This period range was chosen because periods of around 10d are common amongst \TESS{} planet candidates (and reflect the upper end of what is possible for planets detected in single \TESS{} sectors); other period ranges can also be explored with our code. For each size regime, we draw 1000 systems with conditioned planets (i.e. transiting and detectable by \Kepler{}, all within the $P_{\rm cond}$ and $R_{p,\rm cond}$ ranges).

We show scatter plots of $K_{\rm cond}/\sum{K}$ versus $K_{\rm cond}$ in the left--hand panels of Figure \ref{fig:Kcond_Nobs_Rp}. Here, $\sum{K}$ is simply the sum of the $K$ of all the planets in a given system. Thus, $K_{\rm cond}/\sum{K}$ is the fractional contribution of the conditioned planet to the system's total $K$; values of one indicate that the conditioned planet is the only planet in the system. The distribution of points represents the underlying population of systems conditioned on the observed planets as drawn from the \citetalias{2020AJ....160..276H} model. Given the M-R relation adopted in our model, the \Kepler{}--detectable Earth--sized planets tend to have RV semi-amplitudes around $K_{\rm cond} \sim 0.3$ m/s (top panel), while the larger planets have predominantly $K_{\rm cond} \gtrsim 0.3$ m/s (middle and bottom panels). The planets in the top panel with the smallest $K_{\rm cond}$'s require more than $N_{\rm obs} = 300$ observations (red points). The typical planet with $K_{\rm cond}$ close to the single--measurement precision requires roughly $N_{\rm obs} = 50-100$ observations in order to constrain their mass to within 20\% error. This is a strong function of $K_{\rm cond}$ across the $\sigma_{1,\rm obs}$ threshold, but appears to only marginally depend on $K_{\rm cond}/\sum{K}$.

In the right--hand panels of Figure \ref{fig:Kcond_Nobs_Rp}, we plot $N_{\rm obs}$ versus $K_{\rm cond}$ for the same simulations (black points). While the required number of observations clearly decreases with increasing $K_{\rm cond}$, there is a broad distribution of $N_{\rm obs}$ for a given $K_{\rm cond}$, due to the contribution of additional planets. We also fit and plot a linear relation for $\log{N_{\rm obs}}$ as a function of $\log{K_{\rm cond}}$ (the solid red line) for the ideal case in which no other additional planets affect the measurement of $K_{\rm cond}$ (as described in \S\ref{methods:Ideal_case}), equivalent to a power--law of the form:
\begin{equation}
 N_{\rm obs} = N_{\rm obs}(\sigma_{1,\rm obs}) \bigg(\frac{K_{\rm cond}}{\sigma_{1,\rm obs}}\bigg)^\alpha, \label{eq_Nobs_ideal}
\end{equation}
where we find $N_{\rm obs}(\sigma_{1,\rm obs}) \simeq 60$ and $\alpha \simeq -2$. In other words, in the best--case scenario where the transiting planet is the only planet in the system and its period and eccentricity are both known, about 60 observations are needed to measure a signal of $K_{\rm cond} = \sigma_{1,\rm obs}$ to within 20\% error (for planets in the period range of $8-12$ d).
To directly compare with the ideal case, we also plot the cumulative distribution functions (CDFs) of $N_{\rm obs}$ for the black points (solid CDFs) and for the same points as predicted by the ideal case (i.e. vertically matched to the solid red line; dashed CDFs), for various bins in $K_{\rm cond}$ (different colors as labeled). We note that CDFs that do not reach 1 imply that the remaining fraction of the planets require more than 300 observations to measure their $K$'s.

For Earth--sized planets, the required $N_{\rm obs}$ is overall not substantially higher than that of the ideal case (top panel). The distribution of $N_{\rm obs}$ is similar between the solid and dashed CDFs, with the greatest differences seen for the larger $K_{\rm cond}$ bins (e.g. green), although few such planets have $K_{\rm cond} \gtrsim 1$ m/s. Moving to larger planet radii (middle and bottom panels), the differences in the number of observations needed to achieve 20\% accuracy become more considerable, especially as $K_{\rm cond}$ increases to several meters per second, well beyond the single--measurement precision assumed here (0.3 m/s). While any planet with $K_{\rm cond} \gtrsim 1$ m/s would in principle only require $N_{\rm obs} = 5$ observations with $\sigma_{1,\rm obs} = 0.3$ m/s precision if it were the only planet in the system\footnote{In practice, uncertainty in the eccentricity and pericenter direction means that one should not attempt to measure $K$ with so few observations.}, a significant fraction of the time the true minimum number of observations is several times greater. 
For example, planets with 1 m/s $< K_{\rm cond} <$ 2 m/s typically require $10-20$ observations (green CDFs in either panel). Although the larger $K_{\rm cond}$ planets have the greater fractional increase in $N_{\rm obs}$ over the ideal case, the planets with the smallest signals require many more observations and thus the most additional (absolute) number of observations; considering the $K_{\rm cond} < 0.5$ m/s bin (black CDFs in the middle or bottom panels), the median $N_{\rm obs}$ is about 30 for the ideal case but is closer to $50-60$ for planetary systems. This suggests a dual difficulty of measuring the masses of sub--Neptune sized planets with significant gaseous envelopes: dozens of observations with current precision RVs are already typically expected in order to measure their relatively small values of $K_{\rm cond}$, but several dozen more may be necessary in practice due to the presence of other planetary signals.

\subsection{Number of observations as a function of period and radius} \label{results:RV_obs_PR}

\begin{figure*}
\centering
\includegraphics[scale=0.45,trim={0.5cm 0 0.5cm 0.2cm},clip]{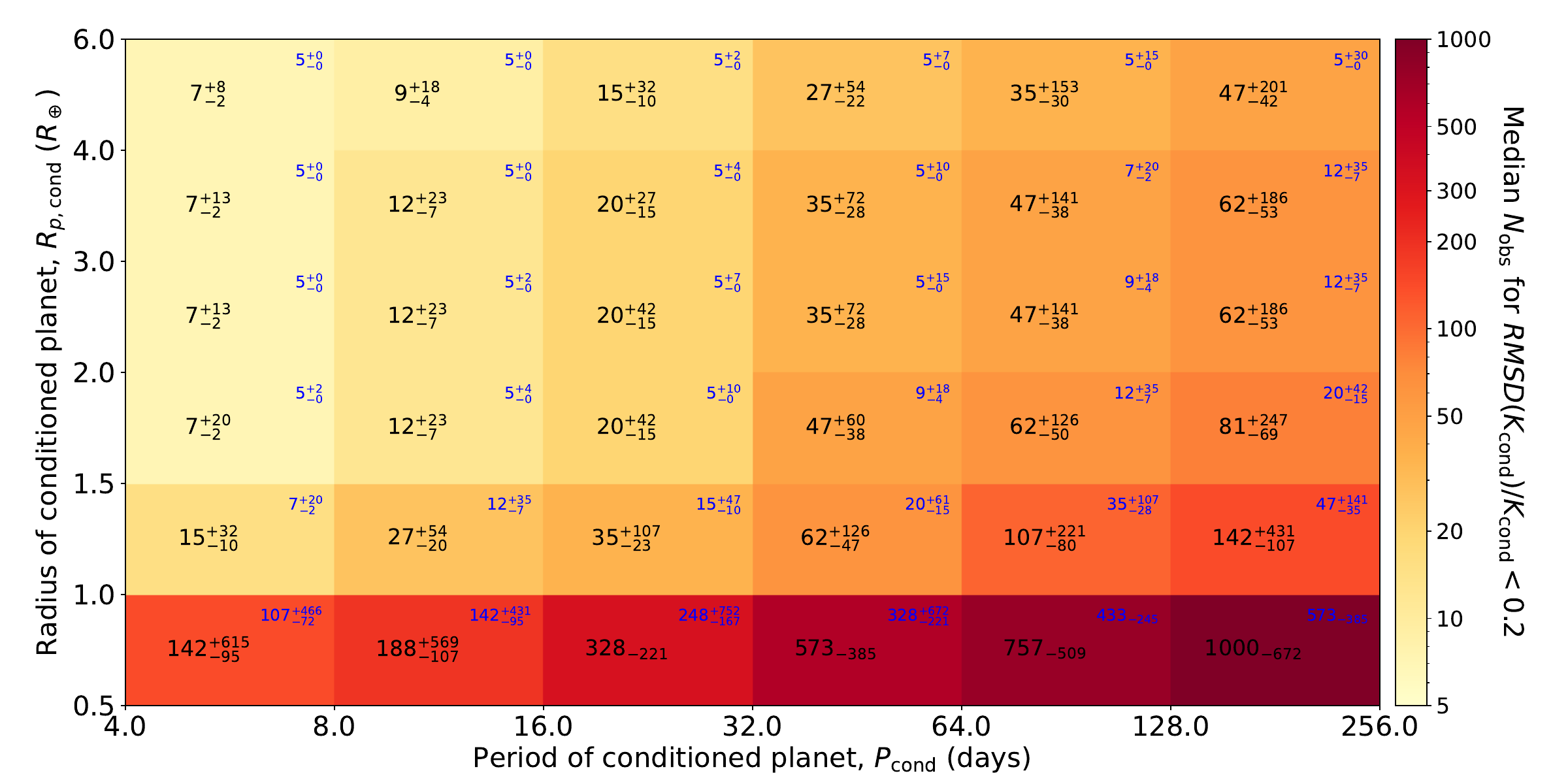}
\caption{Number of observations ($N_{\rm obs}$) required to measure $K_{\rm cond}$ to within 20\% accuracy, for an observed planet for each period--radius bin. The assumed single--measurement RV precision is $\sigma_{1,\rm obs} = 0.3$ m/s. In each cell, the black centered number (also denoted by the color scale) is the median $N_{\rm obs}$, while the blue number in the upper right corner is the number of observations needed in the ideal case (i.e., no additional planets or noise due to stellar variability), $N_{\rm obs, ideal}$. Lower and upper uncertainties, if shown, represent the 16\% and 84\% quantiles, respectively.}
\label{fig:PR_grid_Nobs}
\end{figure*}

In the previous section, we showed how many RV observations are required to measure $K_{\rm cond}$ for planets of several size regimes at orbital periods around $\sim 10$ d. While these relatively short periods are most common for the planets found by \TESS{}, there are a considerable number of transiting planets at longer periods for stars covered by overlapping sectors, which are especially valuable for finding planets in the habitable zones of their host stars. In addition, there is a yield of single--transit events at longer periods for which additional follow--up observations (e.g. with RVs) would be especially useful for confirming their planetary statuses. Finally, there are also continued efforts to measure the masses of the nearest \Kepler{} planet candidates, which have orbits out to $\sim 1$ yr (e.g. the Keck Planet Finder; \citealt{2020SPIE11447E..42G}).

Here, we extend our analyses to test how the typical $N_{\rm obs}$ varies for planets in other regions of period--radius space. In Figure \ref{fig:PR_grid_Nobs}, we show the median and central 68\% $N_{\rm obs}$ on a $P$-$R_p$ grid. For each $P$-$R_p$ bin, we generate 1000 systems conditioned on a transiting and \Kepler{}--detectable planet in that bin and simulate RV observations of those systems, assuming $\sigma_{1,\rm obs} = 0.3$ m/s. We also simulate the ideal case, in which any additional planets in those systems are removed; the central 68\% $N_{\rm obs}$ are shown as blue numbers in the top right corner of each cell. We find that for a given planet radius range, there is a modest increase in the median required $N_{\rm obs}$ with increasing period. It also appears that the values are roughly similar for all bins with $R_p > 1.5 R_\oplus$ for each period range, but increase significantly towards smaller sizes. The median $N_{\rm obs}$ appears to increase ten--fold going from $R_p = 1 - 1.5 R_\oplus$ to $R_p = 0.5 - 1 R_\oplus$, for $P < 32$ d, while the smallest planets with periods longer than that typically require more than 500 observations with 0.3 m/s precision.
Comparing the typical $N_{\rm obs}$ to that of the ideal case (black versus blue numbers in each bin), we emphasize that while the largest planets at the longest periods experience the largest fractional increases, the most substantial increases (in absolute numbers) are for the smallest planets at long periods, often requiring several hundred additional nights of observation.
Finally, we note that the range in $N_{\rm obs}$ (i.e. the denoted uncertainties) also increases substantially for longer periods and for smaller sizes.

\subsection{The effect of RV precision: current and next generation extreme precision RVs} \label{results:RV_precision}

\begin{figure}
\centering
\includegraphics[scale=0.72,trim={0.3cm 0 0 0.5cm},clip]{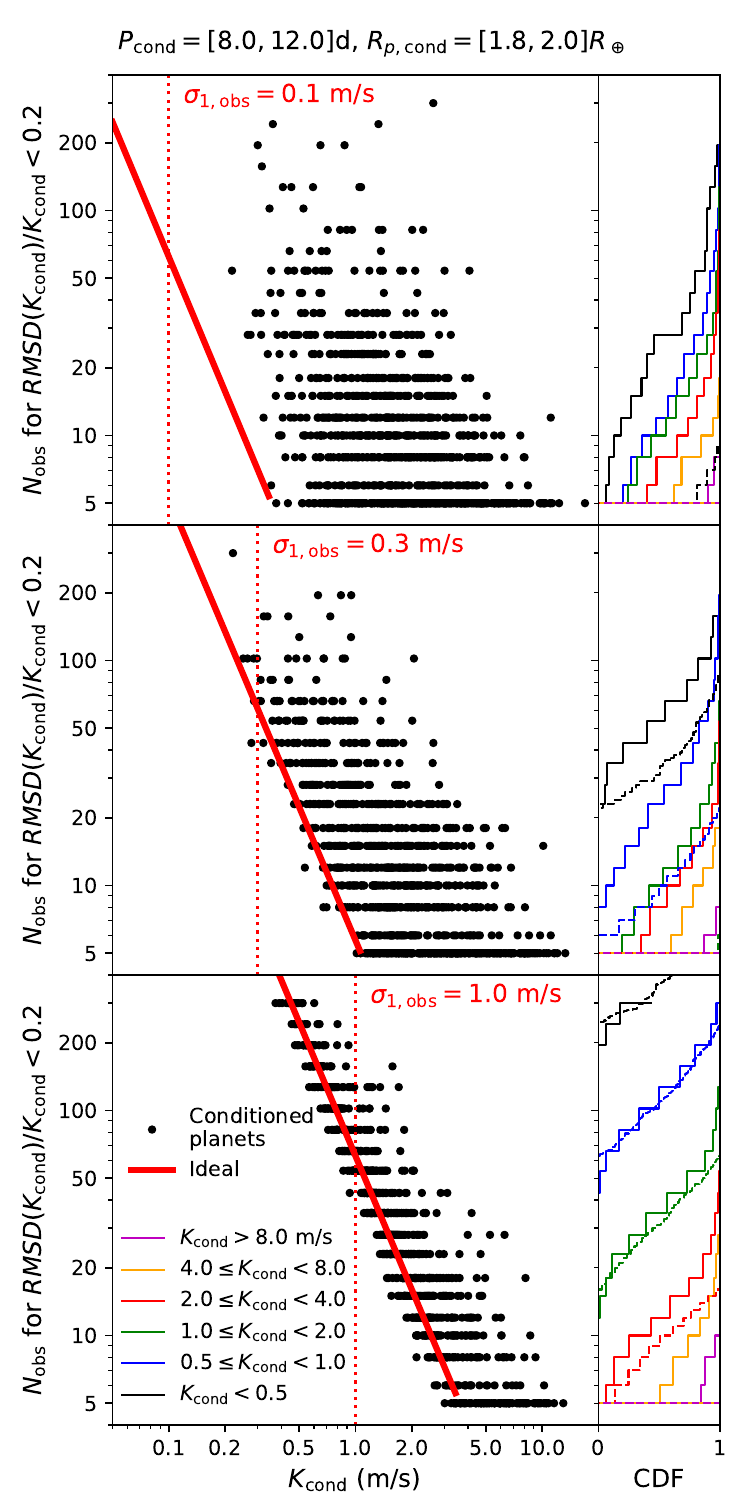} 
\caption{Number of RV observations ($N_{\rm obs}$) required to measure $K_{\rm cond}$ to within 20\% accuracy for conditioned planets in $P_{\rm cond} = [8,12]$ d and $R_{p,\rm cond} = [1.8,2.0] R_\oplus$, for varying single--measurement precisions of: $\sigma_{1,\rm obs} = 0.1$ m/s (\textbf{top panel}), $\sigma_{1,\rm obs} = 0.3$ m/s (\textbf{middle panel}), and $\sigma_{1,\rm obs} = 1.0$ m/s (\textbf{bottom panel}). These panels are analogous to those in the right--hand side of Figure \ref{fig:Kcond_Nobs_Rp}; the black points denote conditioned planets while the red lines show the ideal case. Note that some dashed CDFs overlap and rise to unity at $N_{\rm obs} = 5$ (especially for the upper panels). Here, all panels include a similar distribution of conditioned systems.}
\label{fig:Kcond_Nobs_sigma}
\end{figure}

The current state-of-the-art for EPRVs is pushing $\sim 0.3$ m/s in single--measurement precision (NEID, \citealt{2016SPIE.9908E..7HS}; EXPRES, \citealt{2016SPIE.9908E..6TJ, 2020AJ....159..187P}; and ESPRESSO, \citealt{2021A&A...645A..96P}). A larger number of observatories can achieve a precision in the vicinity of 1 m/s \citep{2016PASP..128f6001F}. Yet, efforts on multiple fronts are continuing to be made towards the goal of achieving 10 cm/s precision, comparable to the amplitude of an Earth--mass planet with a 1 yr orbital period. In this section, we investigate how each of these thresholds in instrumental precision affect our ability to measure $K_{\rm cond}$ for RV followup.

In Figure \ref{fig:Kcond_Nobs_sigma}, we repeat the simulated RV observations for planets conditioned in the range $P_{\rm cond} = 8-12$ d and $R_{p,\rm cond} = 1.8 - 2 R_\oplus$, for three single--measurement RV precision values of $\sigma_{1,\rm obs} = 0.1$, 0.3, and 1 m/s (as denoted and labeled in each panel by the vertical dotted line). In each panel, the distribution of $K_{\rm cond}$ is the same, since the systems are conditioned on the same type of transiting planet. The ideal single--planet case (solid red line) shifts towards the left side (lower values of $K_{\rm cond}$) as $\sigma_{1,\rm obs}$ decreases, but maintains the same fitted relation as given by Equation \ref{eq_Nobs_ideal} (which is normalized to $\sigma_{1,\rm obs}$). As in Figure \ref{fig:Kcond_Nobs_Rp}, we show CDFs of $N_{\rm obs}$ for several bins in $K_{\rm cond}$ on the right side of each panel, with solid lines including the conditioned planets (black points) and dashed lines denoting the expected distributions if these planets were intrinsic singles (black points matched towards the fitted red line).

For an RV precision of 1 m/s (bottom panel of Figure \ref{fig:Kcond_Nobs_sigma}), on first glance the distribution of $N_{\rm obs}$ appears close to the best case scenario; the conditioned planets with $K_{\rm cond}$ near $\sigma_{1,\rm obs}$ (e.g. black, blue, and green CDFs) require a similar number of observations as the ideal case, and only deviate away for larger $K_{\rm cond} \gtrsim 3$ m/s. This trend is similar to the results in the top right panel of Figure \ref{fig:Kcond_Nobs_Rp} (i.e. for $R_{p,\rm cond} = 0.9 - 1.1 R_\oplus$ with $\sigma_{1,\rm obs} = 0.3$ m/s precision).
Since planets with $K$'s close to the measurement precision already require a large number of observations, the presence of additional planetary companions does not lead to a large multiplicative factor in $N_{\rm obs}$. However, the absolute increase in $N_{\rm obs}$ is often substantial, requiring several dozens of additional nights of observation. The situation is quite different as the RV precision improves (relative to $K_{\rm cond}$). Focusing on the $\sigma_{1,\rm obs} = 0.3$ m/s case (middle panel of Figure \ref{fig:Kcond_Nobs_sigma}), there is a significant scatter in the black points above the red line. For example, the median $N_{\rm obs}$ for planets with $K_{\rm cond} < 0.5$ m/s (black CDFs) is $\sim 60$ compared to just $\sim 35$ in the ideal case; for $K_{\rm cond} = 0.5 - 1$ m/s (blue CDFs), the median $N_{\rm obs} \sim 25$ compared to $\sim 10$ if there are no additional planets. Finally, when using a $\sigma_{1,\rm obs} = 0.1$ m/s instrument (top panel), anywhere from $\sim 10 - 100$ observations will be required to accurately measure the planets' RV amplitudes.
In contrast, all the other dashed CDFs (with the exception of the dashed black CDF) shoot up to one at the minimum number of observations we tested, $N_{\rm obs} = 5$.
Thus, recognizing the potential for additional planets to contribute to the RV signal substantially increases the number of observations required to measure the mass of the transiting planet with 20\% accuracy.

We emphasize that these results do not imply that there is nothing to gain from using an extremely precise instrument to follow--up $1.8 - 2 R_\oplus$ planets around 10d orbital periods. The CDFs/median $N_{\rm obs}$ still shift to lower numbers as $\sigma_{1,\rm obs}$ improves, for most $K_{\rm cond}$ bins. Rather, our simulations show that the reduction in $N_{\rm obs}$ is not as rapid as what one would expect in the ideal case, due to the role of additional planets contributing to the measured RVs.

\begin{figure}
\centering
\includegraphics[scale=0.44,trim={0.5cm 0.5cm 0.5cm 0.5cm},clip]{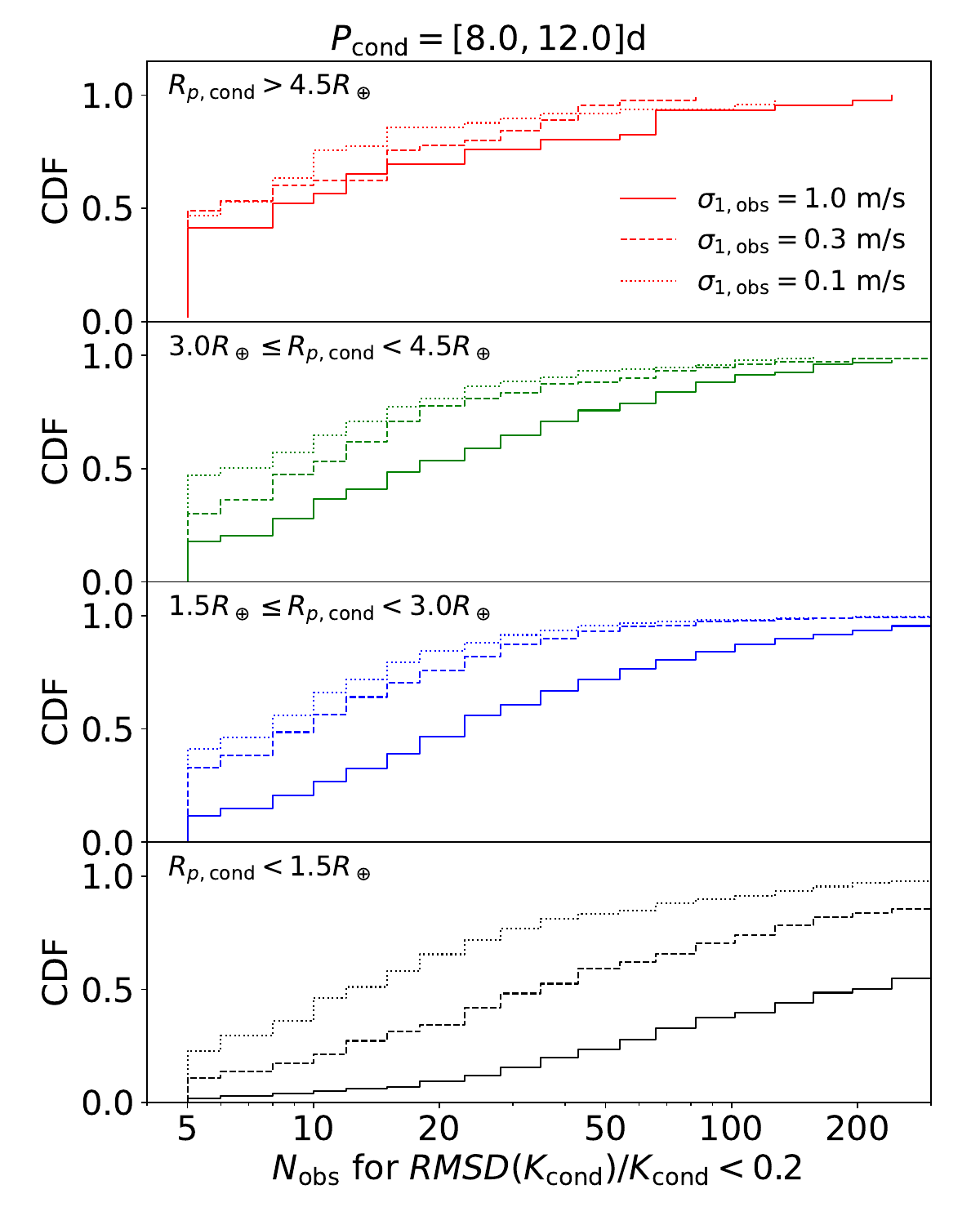} 
\caption{Distribution of the number of RV observations ($N_{\rm obs}$) required to measure $K_{\rm cond}$ to within 20\% accuracy as a function of the conditioned planet's radius, $R_{p,\rm cond}$. All panels include the same range of conditioned periods ($P_{\rm cond} = [8,12]$ d), with ranges in $R_{p,\rm cond}$ as labeled. The required $N_{\rm obs}$ for three different values of single--measurement precision $\sigma_{1,\rm obs}$ are shown.}
\label{fig:Nobs_Rp}
\end{figure}

To further quantify the required observational time with different instrumental precisions as a function of planet size, we plot the cumulative distributions of $N_{\rm obs}$ for $\sigma_{1,\rm obs} = 0.1$, 0.3, and 1 m/s, for several planet radius regimes in Figure \ref{fig:Nobs_Rp}. As with the previous analyses, we choose a period range of $8-12$ d. We find a substantial reduction in the minimum number of observations needed to measure $K_{\rm cond}$ as the measurement precision improves from 1 m/s to 0.3 m/s and to 0.1 m/s, for the smallest radius bin ($R_{p,\rm cond} < 1.5 R_\oplus$). While almost half of these planets require 300 or more observations with 1 m/s precision (to measure their $K$ to better than 20\%), the median $N_{\rm obs}$ reduces to about 30 for observations of 0.3 m/s precision. Improving to 0.1 m/s precision further drops the median $N_{\rm obs}$ to $\sim 10$, and almost all \Kepler{}--observed planets in this size range are measurable with less than 300 observations. The distributions for $1.5 - 3 R_\oplus$ and $3 - 4.5 R_\oplus$ look very similar, with the greatest savings in $N_{\rm obs}$ as $\sigma_{1,\rm obs}$ decreases from 1 m/s to 0.3 m/s (in either case the median $N_{\rm obs}$ reduces from $\sim 20$ to $\sim 8$). For planets larger than $4.5 R_\oplus$, there are only marginal improvements to $N_{\rm obs}$ with $\sigma_{1,\rm obs}$; in any case more than half of the planets can be measured with just 10 or fewer observations.
We caution again that these results are predicated on knowing the orbital period, phase, eccentricity, and argument of pericenter with negligible uncertainty. While the impact of this assumption should be minor for large values of $N_{\rm obs}$, it can be significant for small numbers of observations. Thus, in practice we do not recommend planning for just $\sim~10-30$ observations, even when following up fairly large planets.

Evidently, there is a broad distribution in $N_{\rm obs}$ for any given planet size range and RV precision as shown in Figure \ref{fig:Nobs_Rp}. We note that there are four primary factors contributing to the widths of the distributions in $N_{\rm obs}$ shown here. First, the size of each radius bin is relatively large (larger than the ranges used in Figure \ref{fig:Kcond_Nobs_Rp}); the overall radius distribution roughly follows the \Kepler{}--observed distribution (conditioned on $P_{\rm cond} = 8-12$ d here). Second, there is a scatter in the planet masses given a radius due to our mass--radius (M-R) relation; this shapes the distribution of $K_{\rm cond}$, which strongly predicts the required $N_{\rm obs}$ in the ideal case as discussed in \S\ref{results:RV_obs}. 
There is also scatter in the RV measurement error due to noise, for any given system.
Finally, we have shown that the presence of additional planet companions -- the primary motivation for this study -- can drastically increase the required $N_{\rm obs}$.

We also note that our M-R relation is different for the smallest planet bin than for the larger planets. As described in \S\ref{methods:Models}, our model uses a lognormal distribution in mass (conditioned on radius) centered around an ``Earth--like rocky" model from \citet{2019PNAS..116.9723Z} for planets smaller than $\sim 1.47 R_\oplus$, while the non-parametric \citet{2018ApJ...869....5N} model is adopted for planets above this transition radius. The former includes a scatter about the median that scales with radius, while the latter exhibits a much larger overall scatter (both the median prediction and the scatter are continuous across the transition radius).

\section{Discussion} \label{sec:Discussion}

\subsection{Measuring the mass of a Venus--like planet} \label{discussion:Venus}

\begin{figure*}
\centering
\includegraphics[scale=0.82,trim={0.6cm 0.2cm 0.2cm 0.2cm},clip]{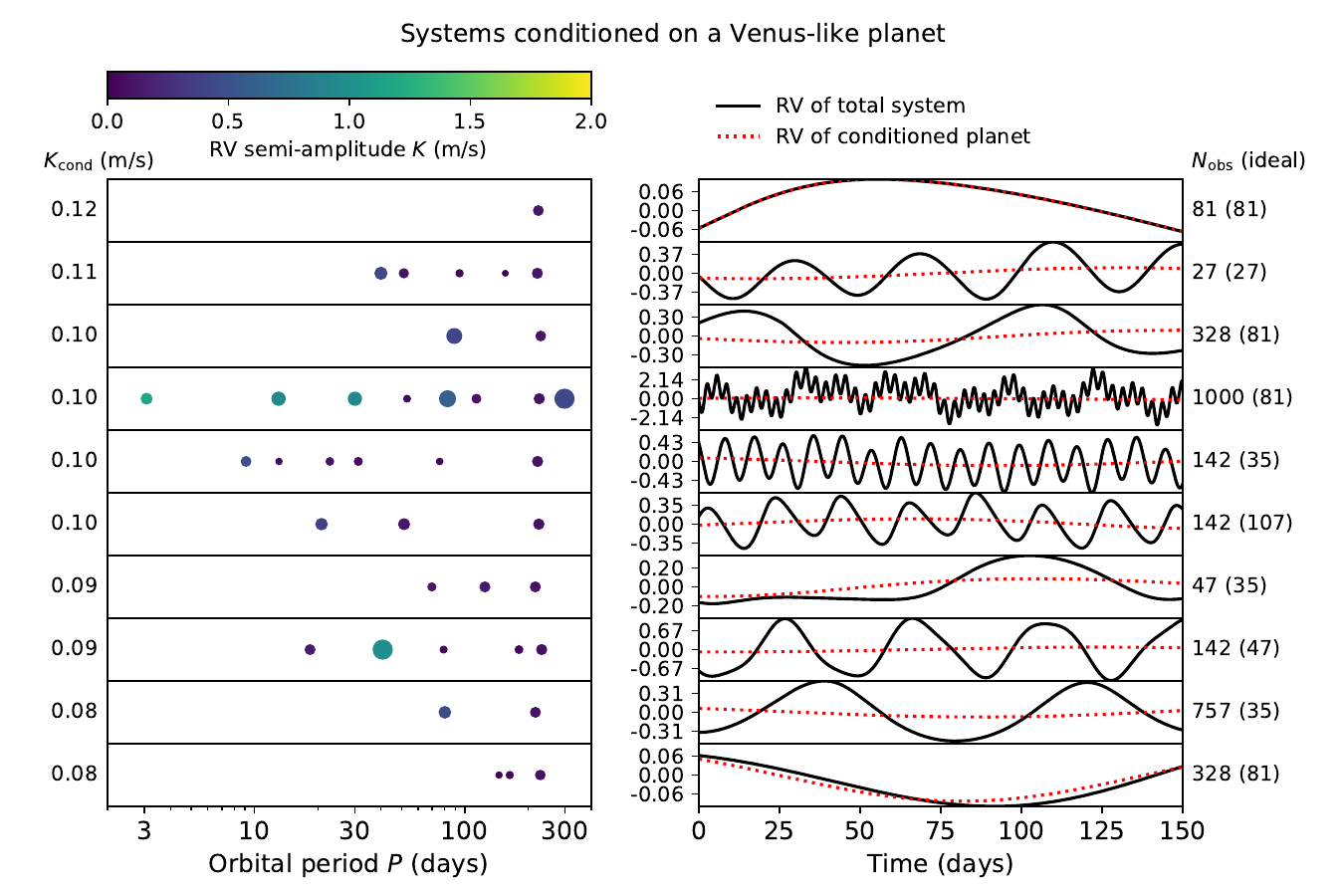}
\caption{Sample of 10 simulated systems with a Venus--like planet. Here, we define ``Venus--like" as any planet with a period, radius, and mass all within $\sim 5\%$ of that of Venus. Similar to Figure \ref{fig:systems_cond_1}, the left--hand column shows a gallery of the systems plotted along the orbital period axis, while the right--hand column shows the true RV time series (the dotted red curves denote the true RV signal of the conditioned planet, i.e. the Venus, only). To the right of each panel, we also list the number of RV observations, $N_{\rm obs}$ (with $\sigma_{1,\rm obs} = 0.1$ m/s), required to measure $K_{\rm cond}$ to within 20\% error assuming a single--planet model with the known orbit of the Venus; the number in parentheses denotes $N_{\rm obs}$ in the ideal case (if the Venus was the only planet in the system, i.e. fitting observations of the dotted red curve).}
\label{fig:systems_cond_Venus}
\end{figure*}

\begin{figure*}
\centering
\includegraphics[scale=0.45,trim={0.8cm 0 0 0.2cm},clip]{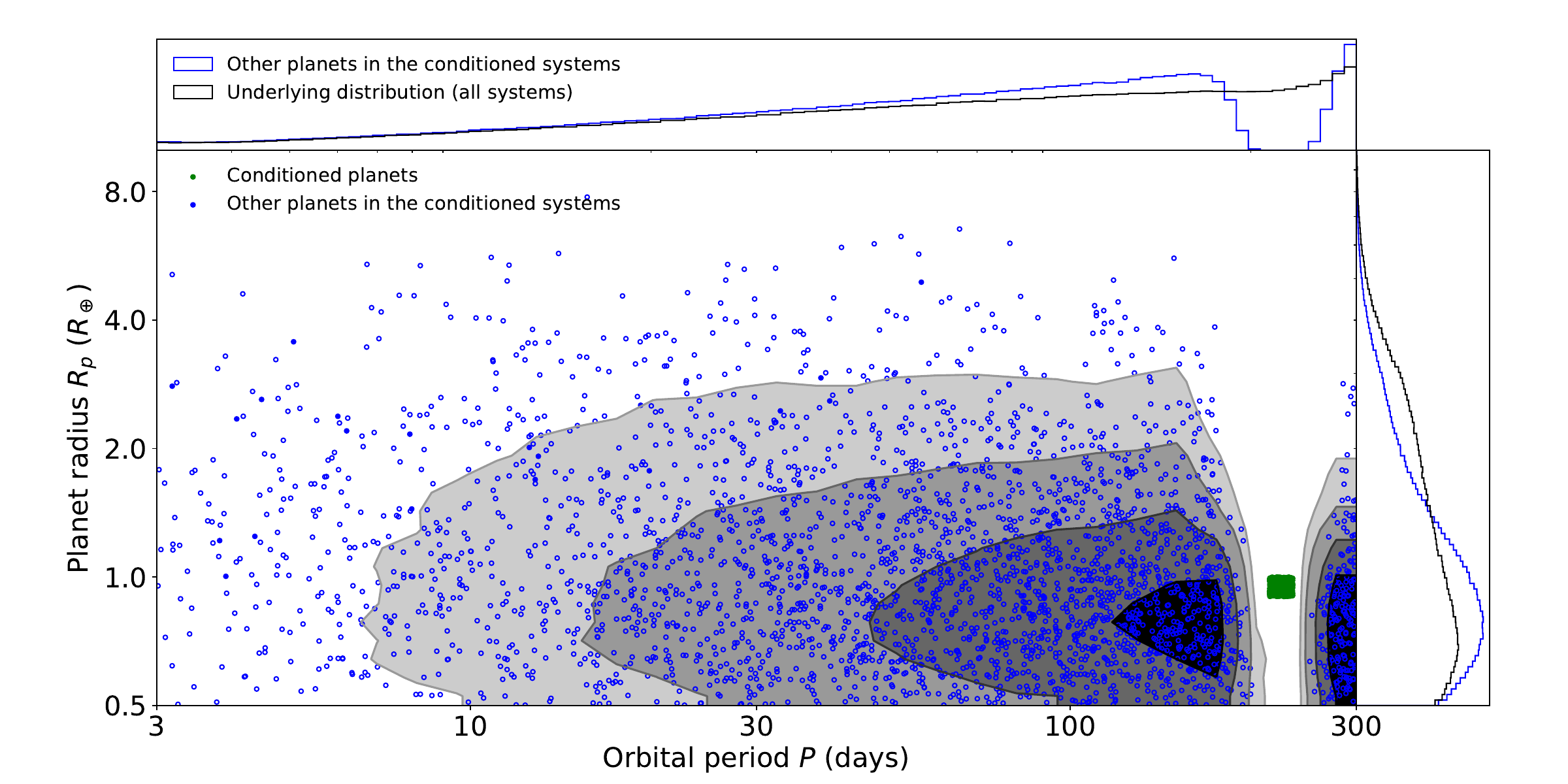}
\includegraphics[scale=0.45,trim={0.8cm 0 0 0.2cm},clip]{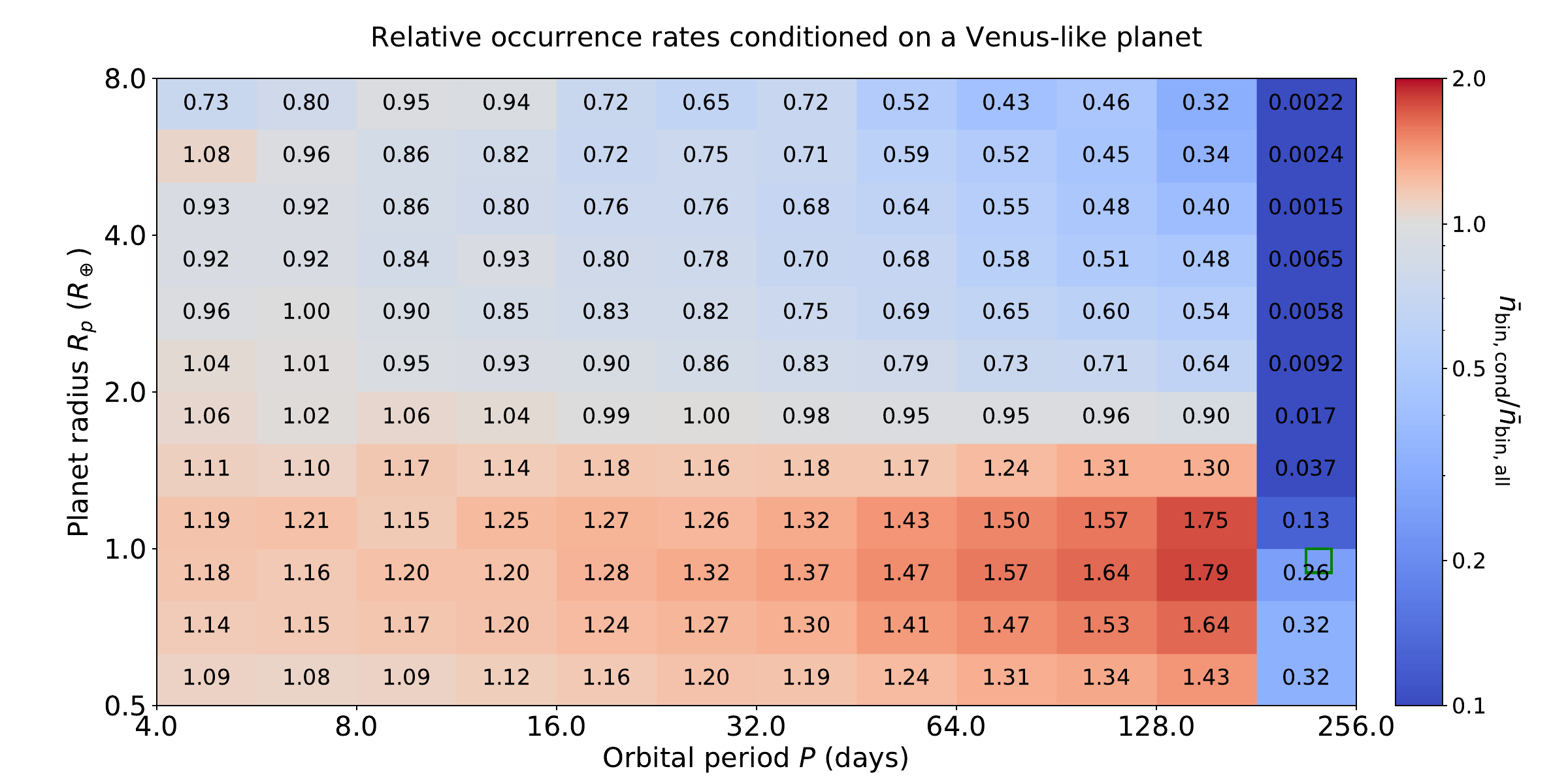}
\caption{Same as Figure \ref{fig:PR_grid_cond_rates}, but conditioned on systems with a Venus--like planet (transiting or not).
As in that figure, green and blue circles (top panel) denote the conditioned (i.e. Venus--like) planets and other planets, respectively, while the gray contours show the distribution of the blue circles. The marginal period and radius distributions are shown by the blue (conditioned systems) and black (all systems) histograms on the top and right panels. The bottom panel reports relative planet occurrence rates on a period-radius grid (mean number of planets per star in the conditioned systems divided by that in all systems, per bin).
A total of $\sim 4.3\times10^5$ conditioned systems are included in the calculations for this figure.
A machine-readable table for the values in the bottom panel is also available online.}
\label{fig:PR_grid_cond_rates_Venus}
\end{figure*}

The \Kepler{} transit survey revealed that planets with sizes similar to that of Earth are abundant in the inner regions of exoplanetary systems. Yet, the mass distribution of these small planets is not as well constrained given the relatively limited and inhomogeneously determined sample of exoplanets with precise masses.
Mass measurements of extrasolar analogs to the terrestrial planets in our own solar system are especially valuable as they may provide key constraints on the compositions and formation processes of these types of planets. 
With a radius of $0.949 R_\oplus$ and a mass of $0.815 M_\oplus$, Venus is only slightly smaller than Earth, but orbits with a shorter period of $224.7$d. 
While the \citetalias{2020AJ....160..276H} model does not include any true Earth--like analogs (it extends to 300d, just within 1 yr orbital periods), we can use the model to explore the prospects for detecting Venus--like exoplanets with future RV surveys.

To investigate how the presence of additional planets affects the mass determination of such planets with RV observations, we draw a large set of simulated planetary systems from our \citetalias{2020AJ....160..276H} model and condition on those that include a ``Venus--like'' planet. Here, we define ``Venus--like'' as any planet with a period, radius, and mass all within approximately 5\% of that of Venus. Similar to Figure \ref{fig:systems_cond_1}, we plot a sample of our simulated systems with a transiting ``Venus--like'' planet (regardless of its detectability by \Kepler{} or any other survey) in Figure \ref{fig:systems_cond_Venus}, with their RV signatures in the right column.
The distribution of planets and their relative occurrence rates in a larger sample of such systems are shown in Figure \ref{fig:PR_grid_cond_rates_Venus} (analogous to Figure \ref{fig:PR_grid_cond_rates}).

\begin{figure}
\centering
\includegraphics[scale=0.42,trim={0.2cm 0.5cm 0 0},clip]{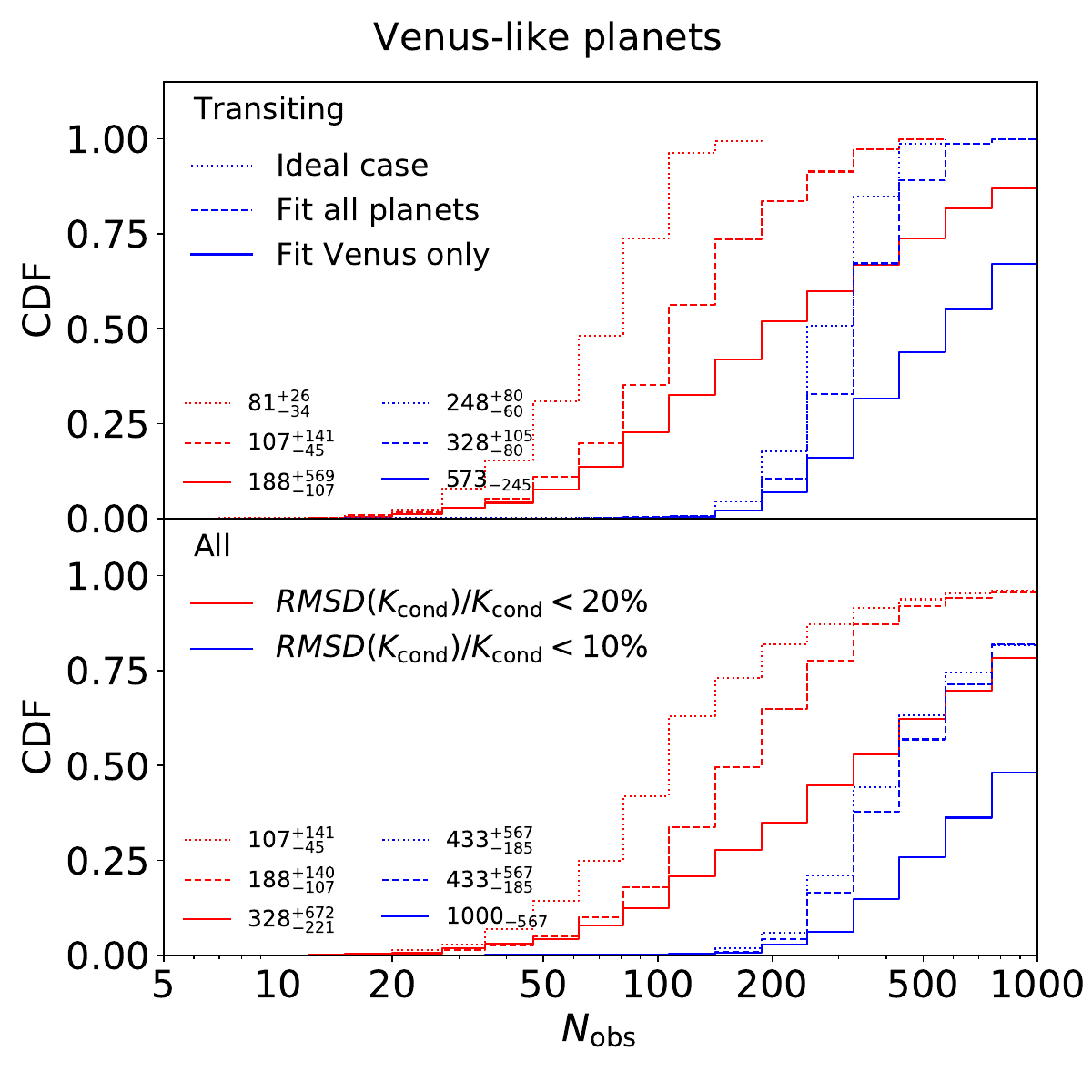}
\caption{Distribution of the number of RV observations ($N_{\rm obs}$) required to measure the $K$ of Venus--like planets, with an instrument of $\sigma_{1,\rm obs} = 0.1$ m/s single--measurement precision. The top panel includes systems with a transiting Venus--like planet (regardless of its detectability by \Kepler{}), while the bottom panel includes all systems with a Venus--like planet (i.e. isotropically distributed). In each panel, solid lines denote the case of fitting the $K$ of the Venus only, dashed lines denote the case of fitting the $K$'s of all planets in the system (assuming known orbits for each planet), and dotted lines show the ideal case (fitting the $K$ of Venus when it is the only planet in the system). Two thresholds for the accuracy in measuring $K_{\rm cond}$ (20\%, red and 10\%, blue) are shown.  In the bottom left corner of each panel, we list the median, 16\%, and 84\% quantiles of $N_{\rm obs}$ for each case.}
\label{fig:Nobs_Venus}
\end{figure}

For each planetary system including a Venus--like planet, we simulate RV observations as in \S\ref{results:RV_obs}. We assume a single--measurement RV precision of $\sigma_{1,\rm obs} = 0.1$ m/s (slightly greater than the RV semi-amplitude of a transiting Venus on a near--circular orbit, $K \simeq 0.086$ m/s, but realistic for a next generation RV spectrograph), no spurious RV signal due to intrinsic stellar variability, and consider surveys with up to $N_{\rm obs} = 10^3$ RV observations. 
We fit for the RV semi-amplitude ($K_{\rm cond}$) of the transiting Venus considering three scenarios: 
(1) the ``ideal'' case where the Venus is the only planet in the system and is fit as such, 
(2) the ``fit Venus only'' case where one still only attempts to fit the $K$ of Venus but unmodeled RV perturbations due to other planets effectively create a source of systematic errors, and
(3) the ``fit all planets'' case where one attempts to measure the $K$'s of all the planets in the system simultaneously, assuming full knowledge of their orbital architectures.
Case (3), while highly optimistic, serves as an intermediate comparison between the other two. In practice, RV observations will often enable the detection of some, but not all, planets in a system. Even then, the orbital periods, phases and eccentricities estimated from RVs may have significant uncertainties.

Figure \ref{fig:Nobs_Venus} (top panel) shows the resulting cumulative distributions of $N_{\rm obs}$ for two requirements on the accuracy of the measured $K$ ($<20\%$ in red and $<10\%$ in blue).
For ease of comparison, the median and 16-84\% quantiles of the distribution for each scenario are listed in the bottom left corner.
This procedure is repeated for a more general case in which the Venus--like planet does not have to be transiting (but rather the orbital planes are distributed isotropically, corresponding to a blind RV survey), in the bottom panel of Figure \ref{fig:Nobs_Venus}.

Focusing first on the transiting Venus--like planets (top panel), we find that in the ideal case, typically $\sim 80$ (250) RV observations with $0.1$ m/s precision are required to measure their $K$ to within 20\% (10\%) error. About $\sim 30\%$ more observations are needed for realistic systems where there are other planets contributing to the RV signature, even when these planets are essentially perfectly accounted for in the RV model (case (3), ``fit all planets'').
An additional $\sim 75\%$ more observations would be necessary to average out the contributions of these other planets when only fitting the Venus (case (2), ``fit Venus only'')! 
Together, one would typically need $\sim 2.3$ times as many observations as what one would naively expect in order to measure the mass of a transiting Venus, when there are an unknown number of planets in the system. 
Our simulations also imply that roughly three times as many observations are needed to improve the accuracy in the measured $K$ from 20\% to 10\%, for our assumed measurement precision. Similar results are seen in the more general case (isotropically distributed Venus--like planets; bottom panel of Figure \ref{fig:Nobs_Venus}), with the distributions of $N_{\rm obs}$ shifting to somewhat larger values. Roughly half of the time, over 1000 observations are needed to measure Venus' $K$ to better than 10\%.
Finally, we caution that even these results are likely optimistic, given that almost all of the planet companions in our models are interior to the orbit of the Venus--like planet (e.g. as seen in Figure \ref{fig:systems_cond_Venus}). 
Additional planets not captured by our models (i.e. with periods $P > 300$ d) could further contribute RV signatures which affect the fitting of the RV data.

In summary, these results imply that in reality, attempting to fit for the mass of a Venus--like planet without accounting for the other planets (i.e. when model misspecification would occur) often requires significantly more (over twice as many) observations than what is usually assumed (i.e., an idealized case) to accurately constrain $K$. 
The complications due to multiple planets will be in addition to other challenges such as intrinsic stellar variability and variable telluric absorption.
Future programs attempting to follow-up and measure the masses of Venus and Earth analogs alike should thus plan for extensive RV monitoring over several years.

\subsection{Which additional planets affect the RV observations the most?} \label{discussion:Which_planets_affect_most}

\begin{figure*}
\centering
\begin{tabular}{cc}
\includegraphics[scale=0.56,trim={0 0.5cm 0 0},clip]{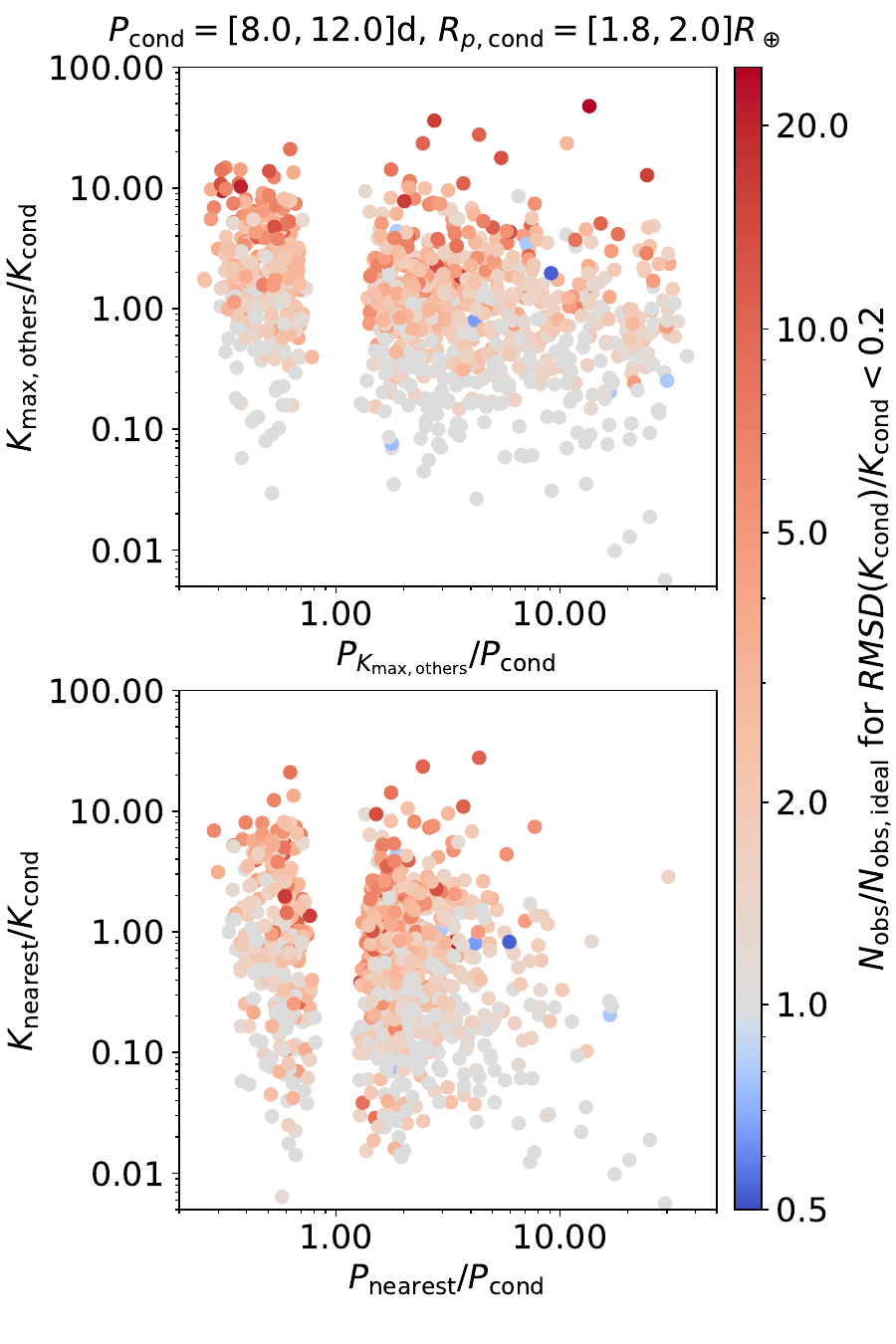} & \includegraphics[scale=0.56,trim={0 0.5cm 0 0},clip]{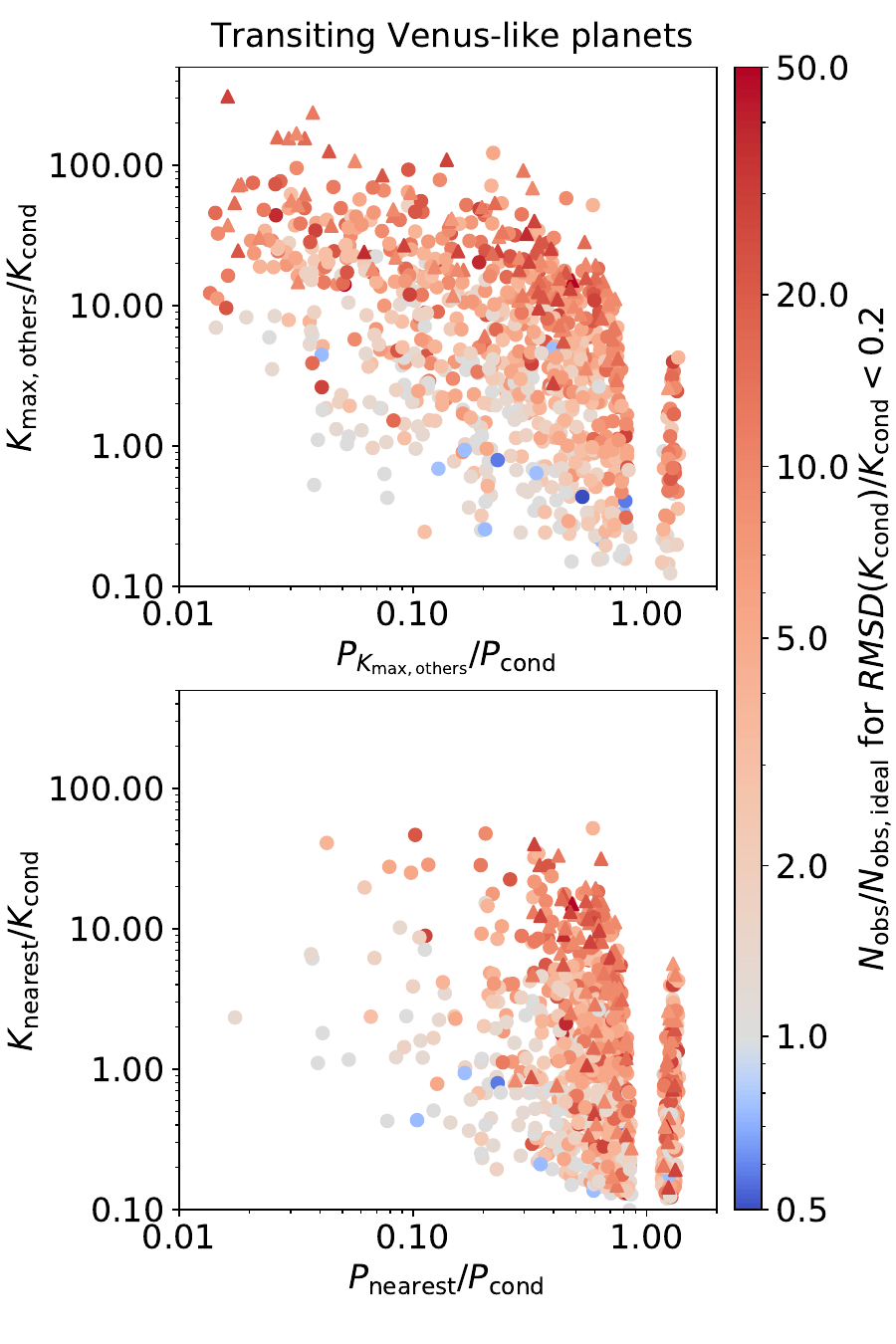}
\end{tabular}
\caption{Scatter plots of RV semi-amplitude ($K$) ratios vs. period ($P$) ratios colored by the ratio of the number of observations needed to measure $K_{\rm cond}$ to within 20\% error relative to the ideal case, $N_{\rm obs}/N_{\rm obs,ideal}$, for \Kepler{}-detectable planets in $P_{\rm cond} = [8,12]$ d and $R_{p,\rm cond} = [1.8,2] R_\oplus$ (\textbf{left-hand panels}; assuming $\sigma_{1,\rm obs} = 0.3$ m/s) and transiting Venus--like planets (\textbf{right-hand panels}; assuming $\sigma_{1,\rm obs} = 0.1$ m/s). In the top panels, $K_{\rm max,others}$ is the semi-amplitude of the planet with the largest $K$ that is not the conditioned planet (and $P_{K_{\rm max,others}}$ is the period of that planet), while in the bottom panels, $K_{\rm nearest}$ is the semi-amplitude of the nearest planet in log-period (whose period is $P_{\rm nearest}$). Both top and bottom panels show the same 1000 conditioned systems. Triangle markers denote lower limits for $N_{\rm obs}/N_{\rm obs,ideal}$ (i.e., the largest attempted $N_{\rm obs}$ is still not enough to achieve better than 20\% accuracy).}
\label{fig:K_P_Nobs_ratios}
\end{figure*}

The results of the RV simulations presented in this paper (\S\ref{results:RV_obs}-\ref{results:RV_precision} and \S\ref{discussion:Venus}) suggest that on average, the presence of additional planets has a substantial impact on inferring the semi-amplitude of the conditioned planet which cannot be ignored. Yet for any given planet, the extent to which the number of RV observations (needed to measure its amplitude accurately) is affected can range from anywhere between a negligible amount to a significant increase, depending on the properties of the unseen planets. Here, we attempt to briefly explore whether there are any discernible patterns in the systems for which the number of RV observations is most affected.

In the top panels of Figure \ref{fig:K_P_Nobs_ratios}, we consider the semi-amplitude and period of the planet with the maximum $K$ in each system (other than the conditioned planet), and plot the scatter of their $K$ ratio versus $P$ ratio relative to the conditioned planet. The points are colored by the ratio $N_{\rm obs}/N_{\rm obs,ideal}$, where $N_{\rm obs}$ is the minimum number of observations necessary to measure $K_{\rm cond}$ assuming the single-planet model and $N_{\rm obs,ideal}$ is the minimum number in the ideal case with all other planets removed (both to better than 20\% error). Two types of conditioned planets are shown: (1) the \Kepler{}-detectable planets with $P_{\rm cond} = [8,12]$ d and $R_{p,\rm cond} = [1.8,2] R_\oplus$ from \S\ref{results:RV_obs} (left--hand panels; RV simulations with $\sigma_{1,\rm obs} = 0.3$ m/s), and (2) the transiting Venus--like planets as defined in \S\ref{discussion:Venus} (right--hand panels; RV simulations with $\sigma_{1,\rm obs} = 0.1$ m/s).

Both cases illustrate that there is a general increase in $N_{\rm obs}/N_{\rm obs,ideal}$ with $K_{\rm max,others}/K_{\rm cond}$; as one might expect, unmodeled planets with $K$ larger than $K_{\rm cond}$ tend to add more systematic noise to RV measurements. When the largest $K$ of the other planets is within a factor of two of the conditioned planet (i.e. $0.5 < K_{\rm max,others}/K_{\rm cond} < 2$), the median $N_{\rm obs}/N_{\rm obs,ideal} \sim 2$. The period dependence is less clear; the trend with $K_{\rm max,others}/K_{\rm cond}$ appears to persist even when the largest $K$ planet has a period ratio $\gtrsim 10$ (both interior and exterior) relative to the conditioned planet. We note that other planets in the same systems not represented in these plots (e.g., planets with $K < K_{\rm max,others}$, which may be closer or further to the conditioned planet and still have considerable $K$'s compared to $K_{\rm cond}$) can also play a role in adding systematic noise to the RV fitting.

Likewise, in the bottom panels of Figure \ref{fig:K_P_Nobs_ratios}, we consider the $K$ ratio and $P$ ratio of the \textit{nearest} planet to the conditioned planet in log--period, in the same sets of systems as in the top panels. While the distributions shift to somewhat smaller $K$ ratios and $P$ ratios closer to unity (by virtue of choosing the nearest planet)\footnote{We note that each conditioned system is represented by a single point in either panel, and some points may be identical in both panels (i.e. when the nearest planet also has the largest $K$ of all the other planets in the system).}, the apparent correlations with $N_{\rm obs}/N_{\rm obs,ideal}$ are qualitatively similar. Again, we reiterate that there may be other planets not represented in these plots which still contribute to the RV signals and thus may also be partly responsible for the increased $N_{\rm obs}$ over the ideal case. Furthermore, it is difficult to disentangle the contribution of the planet with the largest $K$ from that of the nearest planet (when they are distinct).

\subsection{Future improvements} \label{discussion:Improvements}

\subsubsection{Refining the population model}

As models of exoplanetary system architectures improve, new catalogs generated from the updated models can be fed into our framework and directly conditioned on for more refined predictions. While the current model from \citetalias{2020AJ....160..276H} has many features (as summarized in \S\ref{methods:Models}) that make it suitable for the analyses presented in this paper, it also oversimplifies some details about the distribution of planetary systems that should warrant caution and could be modeled more accurately in future iterations. First, while the model was fit to match the marginal distributions of numerous quantities of interest based on the \Kepler{} single- and multi-planet systems, including the marginal distributions of orbital periods and transit depths (which is less sensitive to stellar uncertainties than fitting to the distribution of planet radii directly; see \citealt{2019MNRAS.490.4575H} and \citetalias{2020AJ....160..276H} for the full details of how the models were constrained), it does not capture more complicated features in the observed joint period--radius distribution. One such feature is the observed radius valley, a clear dearth of observed planets around $\sim 1.8 R_\oplus$ that is also a function of orbital period (e.g., \citealt{2017AJ....154..109F, 2018MNRAS.479.4786V}). The \citetalias{2020AJ....160..276H} model adopts a broken power-law for the underlying distribution of planet size scales of each cluster, which cannot naturally reproduce any valley in the radius distribution (underlying or observed).
Future models that reproduce the observed radius valley can potentially be leveraged to more accurately predict the probability of finding another planet above or below the valley in a given system. For example, predictions can be made about the frequency of planets with a gaseous envelope in systems with an inner rocky super--Earth or vice versa, and how they affect the number of RV observations needed to measure the masses of such planets.

\subsubsection{Conditioning on multiple planets in the same system}

The analyses in this paper are limited to conditioning on single planets only, but in theory can also easily be extended to condition on multiple planets. Systems with multiple transiting planets are expected to provide even more powerful predictions for the conditional distribution of planetary companions, due to the strongly correlated nature of planets in the same system and greater constraints for dynamical stability (e.g., gaps between adjacent planets; \citealt{2020AJ....160..107D}). However, the limiting factor of our method when requiring more than one planet within a given set of (narrow) ranges for their properties is computational efficiency, due to the nature of the rejection--sampling--conditioning procedure. We note that while roughly half of all \Kepler{} planet candidates are in multi-transiting planet systems (\citetalias{2020AJ....160..276H}), a smaller fraction of the planets discovered by \TESS{} are in observed multis (168/2241 \TESS{} Objects of Interest from the full prime mission; \citealt{2021ApJS..254...39G}) due to the shorter observing baselines for most targets. Nevertheless, the number of multi-planet systems will grow as more data of previously--monitored targets is obtained, and thus generalizing the code for conditioning on more than one planet is of potentially high scientific value.

\subsubsection{Conditioning on properties of the host stars}

The simulated catalogs can also be trivially conditioned on the host star properties, e.g. by selecting all planetary systems around stars of a given mass or effective temperature. Currently, the only physical dependence of the planets on their host stars in the \citetalias{2020AJ....160..276H} model is in the form of the fraction of stars with planets (within $3-300$ d), which rises towards later type dwarfs. Yet, the detectability of planets in transits is also a function of the properties of their stellar hosts, and thus is an important factor when conditioning on the properties of \Kepler{}--observed planets, as was done in this study. The \textit{SysSim} user interface can be expanded to allow for the input of specific stars in addition to the use of a full \Kepler{} target list, for generating and simulated--observing of planetary systems.

\subsubsection{Including uncertainties in the orbital parameters of the fitted planet}

Future studies could remove the assumption that the orbit is well known. While this is a good assumption for most short-period transiting planets, planets with orbital periods long enough to avoid orbital circularization may have significant eccentricities. For planets detected via direct imaging, it is less clear how well orbits will be characterized independently of RV observations. For planets detected only in RVs, uncertainties in the orbital period, phase, eccentricity and pericenter will make the data analysis more computationally demanding and potentially more challenging. For RV surveys with hundreds of observations and orbital periods of greater than $\sim 100$ days, we anticipate that aliases due to the window function (primarily day-night cycle) will not be as severe a problem as they have been for short-period planets. Nevertheless, there may be similar effects due to aliasing of annual observing constraints. Therefore, it would be prudent to conduct future survey simulations which do not assume the orbits are well known separately from RV observations prior to initiating a large RV campaign.

\subsubsection{Combining the effects of stellar variability}

Finally, future studies should also consider the effects of intrinsic stellar variability in combination with additional planets on the measured radial velocities. This study can be interpreted as assuming either observations of a star with intrinsic stellar variability limited to less than 0.1m/s or observations with sufficient resolution and/or signal-to-noise to recognize the spectral signature of stellar variability and mitigate its effect on the measured radial velocity to be uncorrelated Gaussian noise with scatter less than 0.1m/s. While there has been considerable recent progress in strategies for mitigating the RV perturbations due to stellar activity (e.g., \citealt{2021MNRAS.505.1699C, 2020arXiv201100003D}), it remains unclear if such methods can reach 0.1m/s. If the residual RV contribution from stellar variability is larger and/or correlated in time, then significantly more observations will be required. This paper provides the fractional increase in the number of observations (e.g., Figures \ref{fig:Nobs_Venus} and \ref{fig:K_P_Nobs_ratios}) necessary to provide a first estimate that may be applied to survey simulations that account for the effects of stellar variability. Future research could explore the interaction of multiple planets and stellar variability on the accuracy of exoplanet mass measurements.

\section{Conclusions} \label{sec:Conclusions}

In this study, we used a population model for the distribution of inner planetary systems (\citetalias{2020AJ....160..276H}) to make predictions about the distribution and properties of additional planets in systems conditioned on the period and radius of a known transiting, \Kepler{}--detectable planet (which we refer to as the ``conditioned'' planet). The \citetalias{2020AJ....160..276H} model is suitable for this type of analysis because it was forward modeled to capture numerous population-- and system--level trends seen in the \Kepler{}--observed planetary systems, most notably including the similarly sized planets and clustered periods in the same system, the underlying eccentricity and mutual inclination distributions constrained from AMD stability and the transit duration and duration ratio distributions, and the relative frequency of observed systems at each multiplicity order. We focused on two primary objectives in this paper, to compute the:

\emph{(1) Conditional occurrence of planets given a transiting planet.}
First, we used our tool for conditioning on systems containing a planet in a given range to calculate several statistics of these systems, including the mean number of planets in each system, the fraction of systems where \Kepler{} would have missed a planet interior to the conditioned planet, and the fraction of conditioned planets that do not have the largest $K$ of all the planets in their systems. This is repeated on a grid of period versus radius of the conditioned planet. Some of our key findings are:
\begin{itemize}
 \item The mean number of planets (between $0.5-10 R_\oplus$ and $3-300$ days) in a system with a \Kepler{}--detectable planet (of any size and period in the same range) is $4.50_{-0.45}^{+0.42}$ ($1.75_{-0.14}^{+0.12}$ including only planets with $K > 1$ m/s). This is higher than the mean multiplicity of a planetary system (in the same range) in general, $3.12_{-0.28}^{+0.36}$, since higher multiplicity systems have an increased probability of having at least one detected planet.
 \item A sizable fraction ($0.092_{-0.014}^{+0.015}$) of \Kepler{} planets have at least one missed interior planet with $K > 1$ m/s. This fraction increases with both the period and radius of the conditioned planet. Thus, RV follow-up of transiting planets with sub-m/s precision should anticipate potentially discovering previously unknown planets at periods shorter than that of the transiting planets.
 \item Roughly half of the time, the \Kepler{}--detectable planet is not the planet with the largest $K$ in the system. This fraction increases as $R_{p,\rm cond}$ decreases, since the RV signals of small planets are more likely to be dwarfed by other planets in the same system.
\end{itemize}

\emph{(2) RV follow-up measurements to measure the mass of a transiting planet.}
The results above imply that attempts to accurately measure the mass (i.e. $K$) of a known transiting planet with RV observations will either have to model multiple planetary components, and/or obtain enough observations to average out the systematic ``noise'' from planets other than the transiting one. To quantify these effects, we simulated a series of RV observations on a variety of conditioned systems assuming that the orbit of the transiting planet is known. We simulate how many nightly measurements ($N_{\rm obs}$) are needed to measure $K_{\rm cond}$ to a desired accuracy (typically to within 20\% of the true value), and how it compares to the ``ideal case'' in which no other planets contribute to the RV signal (and any spurious RV signal due to stellar variability has been removed successfully). Our most important results are as follows:
\begin{itemize}
 \item In the ideal case, the minimum $N_{\rm obs}$ required to measure a $K_{\rm cond}$ comparable to the single--measurement precision ($\sigma_{1,\rm obs}$) is about 60, and is proportional to $(K_{\rm cond}/\sigma_{1,\rm obs})^\alpha$ where we find $\alpha \simeq -2$.
 \item For conditioned planets with $K_{\rm cond} \sim \sigma_{1,\rm obs}$ in \Kepler{}--like planetary systems, the typical $N_{\rm obs} \sim 50-100$. Earth--sized planets ($0.9-1.1 R_\oplus$) around periods of 10 days typically have $K_{\rm cond}$ around $\sim 0.3$ m/s, comparable to the $\sigma_{1,\rm obs}$ of the most precise RV instruments currently being tested.
 \item For conditioned planets with $K_{\rm cond}$ well above $\sigma_{1,\rm obs}$, the required $N_{\rm obs}$ can still be significantly (several times) higher than the number needed in the ideal case due to other planets.
 \item Unsurprisingly, improvements in $\sigma_{1,\rm obs}$ lead to the greatest reductions in $N_{\rm obs}$ for the smallest planets. For planets with $R_{p,\rm cond} < 1.5 R_\oplus$ around 10-day periods, improving $\sigma_{1,\rm obs}$ from 1 m/s to 0.3 m/s reduces the median $N_{\rm obs}$ from $\sim 300$ to $\sim 30$.
 \item We test the prospects of measuring Venus--like planets (defined as those having a period, radius, and mass within $\sim 5\%$ of that of Venus) with next generation RV instruments achieving $\sigma_{1,\rm obs} = 10$ cm/s, and find that $N_{\rm obs} \sim 200$ observations are typically needed to measure their $K$ to within 20\% accuracy (again assuming the planet transits, its orbit is known, and there are no spurious RV signals due to stellar variability). This is $\sim 2.3$ times as many observations as the typical (median) number in the ideal case. Roughly three times as many observations are needed to improve the accuracy in the inferred $K$ from 20\% to 10\%.
 \item If one knew of all the planets (and their precise orbits) in a system in addition to the Venus--like planet and attempted to fit their $K$'s simultaneously, one could typically expect to save roughly $\sim 75\%$ in the number of observations (relative to treating the RV signals of undetected planets as noise).
 \item In the above case (fitting the RV semi-amplitudes of all the planets with exactly known orbits), one still typically requires $\sim 1.3$ times as many as in an idealized case where the Venus--like planet is the only planet in the system.
\end{itemize}

In summary, while the detection of new planets in known systems will provide us with a more complete picture of planetary system architectures and further constrain population models, these planets also pose a unique challenge in the characterization of the known planets with RV observations. Overall, the impact of additional planets in the RV data of known systems cannot be easily ignored, but instead should be taken into account when planning RV follow-up surveys for mass determination.

This work also informs the question of combining RV and transit surveys to better understand the underlying population. It highlights the challenges of discovering and characterizing tightly-spaced low-mass planets with amplitudes comparable to the observation uncertainties (see also Ragozzine et al., in prep.). More work is needed to self-consistently incorporate planets discovered by RVs and their masses into population models while accounting for selection biases. In principle, a joint analysis of multiple data sets from RV and transit observations using a hierarchical model could be performed. In practice, astronomers will have to carefully select their sample in both modelling and further observations (e.g., the Magellan--\TESS{} Survey; \citealt{2020arXiv201111560T}). Other methods of measuring planetary masses (such as transit timing variations characterized using a photodynamical model) would avoid such challenges while measuring many of the same properties of the underlying population.

While the \TESS{} mission is nearing its primary goal of determining the masses of 50 small planets ($R_p < 4 R_\oplus$) alerted from its prime mission, the majority of its $\sim 2250$ \TESS{} Objects of Interest (TOIs) still do not have any mass measurements \citep{2021ApJS..254...39G}. Continued efforts to confirm and measure the masses of these \TESS{} planet candidates with RV follow-up observations could benefit from the conditioning of population models, as demonstrated in this paper. For example, ground-based telescope time can be more efficiently allocated given the considerations for additional planetary signals to that of the transiting planet in RV data. Our model allows observers to compute a more realistic estimate for the number of observations likely required to achieve a certain accuracy in measuring the planet mass, and thus can be used to optimize the number and selection of targets for RV follow-up with a given amount of telescope time. Information about likely orbital periods could also help inform the time-span and cadence of these observations. Finally, the results of our study imply that many more as-yet undiscovered planets reside in the inner regions of these known systems. For example, planet occurrence is likely enhanced at small period ratios to a known transiting planet (barring regions of orbital instability) compared to a random FGK star (e.g. Figure \ref{fig:PR_grid_cond_rates}). In addition to the prospects of confirming TOIs as bona fide planets and accurately measuring their masses, the prioritization of \TESS{} targets for follow-up observations could also consider the conditional probability of additional detectable planets in the same systems.

The \textit{SysSim} code with the functionality for simulating catalogs with systems conditioned on a given planet is publicly available at \url{https://github.com/ExoJulia/SysSimExClusters}, along with full, pre-generated \Kepler{}--like catalogs as described in \citetalias{2020AJ....160..276H}. It can be used to perform analyses more specific to a given planet than the results presented in this paper, for example as applied to individual planet candidates discovered from the \TESS{} primary and extended missions.

\acknowledgments

We thank the entire \Kepler{} team for years of work leading to a successful mission and data products critical to this study.  
We acknowledge many valuable contributions with members of the \Kepler{} Science Team's working groups on multiple body systems, transit timing variations, and completeness working groups.  
We thank an anonymous referee for useful suggestions for improving the analysis and manuscript.
We thank Keir Ashby, Danley Hsu, and Robert Morehead for contributions to the broader SysSim project.

M.Y.H. acknowledges the support of the Natural Sciences and Engineering Research Council of Canada (NSERC), funding reference number PGSD3 - 516712 - 2018.
E.B.F. and D.R. acknowledge support from the Exoplanet Research Program grant \# NNX15AE21G. 
This work was supported by a grant from the Simons Foundation/SFARI (675601, E.B.F.).
E.B.F. acknowledges the support of the Ambrose Monell Foundation and the Institute for Advanced Study.  
M.Y.H. and E.B.F. acknowledge support from the Penn State Eberly College of Science and Department of Astronomy \& Astrophysics, the Center for Exoplanets and Habitable Worlds, and the Center for Astrostatistics.  
E.B.F. acknowledges support and collaborative scholarly discussions during  residency at the Research Group on Big Data and Planets at the Israel Institute for Advanced Studies.  
The citations in this paper have made use of NASA's Astrophysics Data System Bibliographic Services.  
This research has made use of the NASA Exoplanet Archive, which is operated by the California Institute of Technology, under contract with the National Aeronautics and Space Administration under the Exoplanet Exploration Program.
This work made use of the stellar catalog from \citet{2019AJ....158..109H} and thus indirectly the gaia-kepler.fun crossmatch database created by Megan Bedell.
We acknowledge the Institute for Computational and Data Sciences (\url{http://icds.psu.edu/}) at The Pennsylvania State University, including the CyberLAMP cluster supported by NSF grant MRI-1626251, for providing advanced computing resources and services that have contributed to the research results reported in this paper.

\software{ExoplanetsSysSim \citep{eric_ford_2018_1205172},
          SysSimData \citep{eric_ford_2019_3255313},
          Numpy \citep{2011CSE....13b..22V},
          Matplotlib \citep{2007CSE.....9...90H}
          }


\bibliographystyle{aasjournal}
\bibliography{main}





\end{document}